\documentclass[proof,dvipdfmx]{pasj01}

\Received{2014 November 25}
\Accepted{2015 April 6}

\newcommand{\oppardif}[1]{\frac{\partial}{\partial #1}}
\newcommand{\pardif}[2]{\frac{\partial #1}{\partial #2}}

\newcommand{\unit}[1]{\rm ~ #1}
\newcommand{\iso}[2]{{}^{#1}{\rm #2}}

\begin{document}

\title{
Revealing progenitors of type Ia supernovae from their  light curves and spectra
}
\author{
  Masamichi Kutsuna \altaffilmark{1,2,3},
            and 
              Toshikazu Shigeyama \altaffilmark{1}
}
\altaffiltext{1}{
  Research Center for the Early Universe,
           Graduate School of Science, University of Tokyo,
           Bunkyo-ku, Tokyo 113-0033, Japan.
}
\altaffiltext{2}{
  Department of Astronomy,
             Graduate School of Science, University of Tokyo,
             Bunkyo-ku, Tokyo 113-0033, Japan.
}
\altaffiltext{3}{current address: 
  Yasuda gakuen senior high school,
             Sumida-ku, Tokyo 130-8615, Japan.
}
\email{shigeyama@resceu.s.u-tokyo.ac.jp}

\maketitle
%\shortauthors{KUTSUNA \& SHIGEYAMA}

\begin{abstract}
In the single degenerate (SD) scenario of type Ia supernovae (SNe Ia),
the collision of the ejecta  with its companion results in stripping  hydrogen rich matter  from the companion star. This hydrogen rich matter might leave its trace in the light curves and/or spectra.
In this paper,
we perform radiation hydrodynamical simulations of this collision for three binary systems.
As a result, we find that the emission from the shock-heated region is not as strong as in the previous study. This weak emission, however, may be a result of our underestimate of the coupling between the gas and radiation in the shock interaction.
Therefore, though our results suggest that the observed early light curves of SNe Ia can not rule out binary systems with a short separation as the progenitor system,  more elaborate numerical studies will be needed to reach a fair conclusion.
Alternatively,  our results indicate that the feature observed in the early phase of a recent type Ia SN 2014J might result from interaction of the ejecta with a red giant companion star.
We also discuss  the dependence of spectral features of H$\alpha$ and Si II absorption lines  on viewing angles and investigate whether they can constrain the event rate of the SD progenitor.
\end{abstract}

\section{Introduction}
Type Ia supernova (SN Ia) is thought to originate from an accreting carbon-oxygen white dwarf (WD) in a binary system \citep{1973ApJ...186.1007W,1984ApJS...54..335I,1984ApJ...286..644N,1984ApJ...277..355W}.
However, it is uncertain what is the companion star supplying its mass.
There are two scenarios to explain SNe Ia.
One is the single degenerate (SD) scenario, in which the companion star is a main sequence or a red giant star \citep{1984ApJS...54..335I,1984ApJ...286..644N}.
The envelope of the companion star overflows the Roche lobe, and transfers to the surface of the WD.
The other is the double degenerate (DD) scenario, in which two WDs merge and end up with explosion \citep{1984ApJS...54..335I}.
To distinguish the two scenarios, we need to find observable differences between them.

 In the SD scenario, SN ejecta collide with its companion. 
\citet{2010ApJ...708.1025K} simulated this collision
and found
 that high energy photons are emitted from the shock-heated region into a certain limited solid angle
and that the emission becomes prominent before the peak of the light curve due to $\iso{56}{Co}$ decay.
This emission is strong especially in a WD + red giant system.
\citet{2010ApJ...722.1691H} investigated early phases of observed light curves gathered by the Sloan Digital Sky Survey (SDSS)  to see whether they have this prompt emission.
By comparing its early light curve with the observations,
they found that there is no obvious emission feature in the light curves,
which constrained the event rate of each model.
As a result, a progenitor system with a main sequence companion more massive than $6M_\odot$ or a red giant is ruled out as the primary source of SNe Ia.

\citet{2010ApJ...708.1025K} assumed the LTE condition and further that whole matter is composed of radiation dominated gases with $\gamma = 4/3$.
We suspect that they overestimated the radiation temperature of the emission regions. In fact, based on their detailed calculations of spectra from the same hydrodynamical models presented here, \citet{2014ApJ...794...37M} argued that hydrogen rich matter filling in the hole  excavated by the companion star  prevents the photosphere from quickly receding to the Ni-rich region and makes the SN redder than expected by the previous study \citep{2004ApJ...610..876K}. %Maeda+ 2014
In addition,  \citet{2010ApJ...708.1025K} focused only on the impacts on light curves.
We investigate also the influence on spectra.
If the material stripped from the companion does not spread in a wide solid angle,
hydrogen lines from the companion might appear in the spectrum, depending on the line of sight. 

\citet{2007ApJ...670.1275L} investigated nebula spectra of some SNe Ia and found the upper limit of $0.01M_\odot$ for solar abundance material.
Several authors \citep{1975ApJ...200..145W, 2000ApJS..128..615M, 2008AA...489..943P}
 calculated the mass of the material stripped from the companion star by the collision.
\citet{2000ApJS..128..615M} performed hydrodynamical simulations to estimate the stripped mass.
According to their results, the stripped masses are $0.15M_\odot$ for the $1M_\odot$ main sequence companion, $0.53M_\odot$ for the $1M_\odot$ red giant companion.
On the other hand, there is no spectral feature of the stripped material in observed spectra.
Recently, \citet{2008AA...489..943P} adopted a more realistic profile for a main sequence companion from a binary evolution theory
and found that the stripped mass decreases to $0.01-0.02M_\odot$.
 \citep{2004ApJ...610..876K} investigated signatures of hydrogen lines in the  phase later than the maximum light when the shock heating due to the collision has been already consumed by adiabatic expansion.

In this study, we relax the assumption of the instantaneous coupling between gas and radiation
to examine whether or not the decoupling of gas and radiation changes the influence on light curves found by \citet{2010ApJ...708.1025K}.
In addition, we investigate whether hydrogen in the stripped material leaves its trace in spectra especially before maximum light. In later phases, \citet{2014ApJ...794...37M} already studied spectra for the same model presented here.
At first, we simulated the collision of a SN Ia with its companion with a radiation hydrodynamical code.
Secondly, we calculated spectra by ray-tracing snapshots of the distributions of density, temperatures, and velocity obtained from the simulation.
Section 2 describes our models.
Section 3 and 4 present the numerical methods for the simulation and ray-tracing.
In section 5, we describe our results.
Section 6 concludes this study.

\section{Models}
As progenitor systems, we consider three models named MS, RGa, and RGb (Table.\ref{model.tbl}).
%One is the case where the companion star is a main sequence star (called model MS).
Model MS is a close binary system with a main sequence companion.
We use a polytropic star with the index $n = 3$ to model the structure of the companion star.
We assume the companion star composed of a solar abundance material.
Its mass $M_2$ is equal to $1\,M_\odot$.
The binary separation $A$ is equal to $3\times10^{11}\unit{cm}$.
This model is the same as the $1\,M_\odot$ main sequence model in \citet{2010ApJ...708.1025K}.

Model RGa is a close binary system with a red giant companion.
In this model,
the companion star consists of the helium core and the hydrogen-rich envelope with the solar abundance.
Its core mass and total mass $M_2$ are equal to $0.4M_\odot$ and $1M_\odot$, respectively.
The binary separation $A$ is equal to $2\times10^{13}\unit{cm}$.
We assume a fully convective envelope with $n=3/2$.
In the hydrodynamical simulation, the core is treated as a point source of gravity.
This model is the same as the $1\,M_\odot$ red giant model in \citet{2010ApJ...708.1025K}.
Model RGb is the same as model RGa but with a longer separation,
i.e., $A=3\times10^{13}\unit{cm}$.

The radii $R_2$ of the companion stars in the above models are obtained by an empirical law for close binary systems \citep{1971ARA&A...9..183P} assuming that the companion stars fill the Roche lobe.
\begin{equation}
  \frac{R_2}{A} = 0.38 + 0.2 \log \frac{M_2}{M_\mathrm{WD}}.
  \label{method-separation}
\end{equation}
Here,
$M_\mathrm{WD}$ is the mass of the progenitor WD, which is equal to $1.38M_\odot$.
Thus $R_2 = 0.35A$.

\begin{table}
  \caption{Models in our calculation.}
  \label{model.tbl}
  \begin{center}
    \begin{tabular}{lccc}
      \hline
      \hline
      Model & $M_2$ ($M_\odot$) & $R_2$ ($10^{13}\unit{cm}$) & $A$ ($10^{13}\unit{cm}$)
      \\
      \hline
      MS  & 1 & 0.01 & 0.03
      \\
      RGa & 1 & 0.7 & 2
      \\
      RGb & 1 & 1 & 3
      \\
      \hline
    \end{tabular}
  \end{center}
\end{table}

\section{Simulation}
We simulate the collision between a SN Ia and its companion.
This simulation is performed with a radiation hydrodynamical code in the axisymmetric spherical coordinate system $(r,\theta)$.
We ignore the orbital motion of the progenitor binary.
The orbital velocities are calculated as
$190 \unit{km/sec}$ for model MS,
$23 \unit{km/sec}$ for model RGa,
and $19\unit{km/sec}$ for model RGb.
The material expanding much faster than these velocities is not affected by the orbital motion.
Since the material slower than these velocities is located inside the photosphere over the relevant period,
we do not observe the influence of the orbital motion.

The basic equations governing hydrodynamics in an Eulerian coordinate system are written as follows.
\begin{equation}
  \pardif{\rho}{t} + \nabla\cdot (\rho \mathbf{v}) = 0,
  \label{hydro-mass}
\end{equation}
\begin{equation}
  \oppardif{t}(\rho\mathbf{v}) + \nabla\cdot (\rho\mathbf{v}\mathbf{v})
  = - \nabla p + \mathbf{f}_\mathrm{ext},
\end{equation}
\begin{equation}
  \oppardif{t} \left[\rho \left(\frac{v^2}{2} + e\right)\right]
  + \nabla\cdot \left[\rho\mathbf{v}\left(\frac{v^2}{2} + h\right)\right]
  = \mathbf{v} \cdot \mathbf{f}_\mathrm{ext} + \mathcal{S},
  \label{hydro-energy}
\end{equation}
where $t$ is the time,
$\rho$  the density,
$\mathbf{v}$  the fluid velocity,
$p$  the gas pressure,
$e$  the specific internal energy of gas,
$h = e + p$  the specific enthalpy,
$\mathbf{f}_\mathrm{ext}$  the external force,
and $\mathcal{S}$ is the energy generation rate from heat sources.
For ideal gases, $p$ and $e$ are expressed by the equation of state with the adiabatic index $\gamma$,
\begin{eqnarray}
  p &=& \frac{\rho k T}{\mu m_p},
  \\
  e &=& \frac{\gamma}{\gamma - 1}\frac{\rho k T}{\mu m_p},
\end{eqnarray}
where
$k$ is the Boltzmann constant,
$m_p$  the proton mass,
$\mu$  the mean molecular weight,
and $T$ is the temperature.
In addition, the density of each isotope can be calculated by equations
\begin{equation}
  \oppardif{t}(X_i\rho) + \nabla\cdot (X_i \rho \mathbf{v}) = D_i,
  \label{hydro-isotope}
\end{equation}
where $X_i$ and $D_i$ are the mass abundance and the production/decay rate of the $i$-th isotope.
Simultaneously, we solve the bolometric radiation energy equation,
assuming the flux limited diffusion for the radiation flux $\mathbf{F}$.
\begin{equation}
  \pardif{E}{t}
  + \nabla\cdot \left(\mathbf{F}+ \frac{4}{3}\mathbf{v}E\right)
  = - \mathbf{v}\cdot\frac{\chi}{c}\mathbf{F} + c\alpha (a T^4 - E),
\end{equation}
\begin{equation}
  \mathbf{F} = \frac{- c \nabla E}{3\chi + |\nabla E|/E},
  \label{rhd-end-eq}
\end{equation}
where
$c$ is the speed of light in vacuum,
$a$  the radiation constant,
$\chi = \alpha+\sigma$,
$\alpha$  the absorption coefficient,
 $\sigma$  the scattering coefficient, and
$E$ is the radiation energy density. The flux limited diffusion approximation has been widely used in multi-dimensional radiation-hydrodynamics simulations \citep[ for example]{2009PASJ...61L...7O}.
Considering the interaction between gas and radiation, $\mathbf{f}_\mathrm{ext}$ and $\mathcal{S}$ are written as,
\begin{equation}
  \mathbf{f}_\mathrm{ext} = \rho \mathbf{g} + \frac{\chi}{c}\mathbf{F},
\end{equation}
\begin{equation}
  \mathcal{S} = c\alpha (E - a T^4) + \epsilon,
  \label{eqrhd-end}
\end{equation}
where
$\mathbf{g}$ is the gravitational field,
and
$\epsilon$ is the heating rate by nuclear reactions.

To solve the above system of equations,
we separately integrate each term with respect to time
according to the operator splitting method with first order accuracy in time.
We integrate the advection terms with the Harten-Lax-van Leer-Contact (HLLC) scheme \citep{1994ShWav...4...25T}.
In addition,
we use the Monotonic Upstream-Centered Scheme for Conservation Laws (MUSCL) scheme \citep{1976cppa.conf...E1V}
to obtain solutions with the second order accuracy in space.
We solve the radiative diffusion and the thermalization processes with an implicit method.

\subsection{Initial condition}
\begin{figure*}
  \begin{center}
   \includegraphics{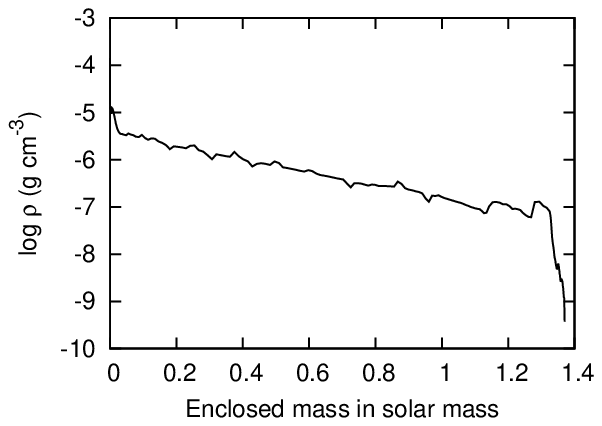}
    \includegraphics{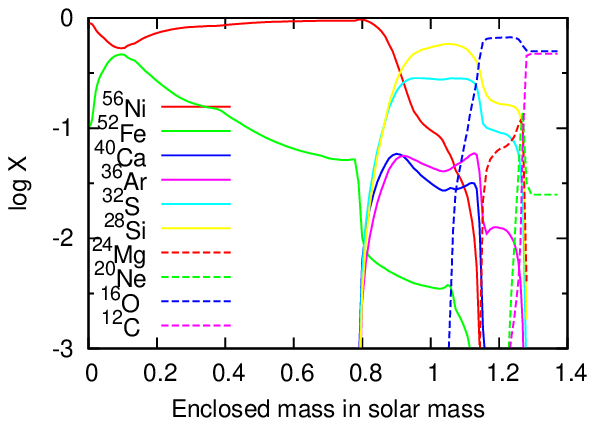}
  \end{center}
    \caption{Density and abundance structure of the ejecta at 0.1 days after the explosion.}
    \label{study-w7.eps}
\end{figure*}

\begin{figure*}
  \begin{center}
  \includegraphics[width=0.3\linewidth]{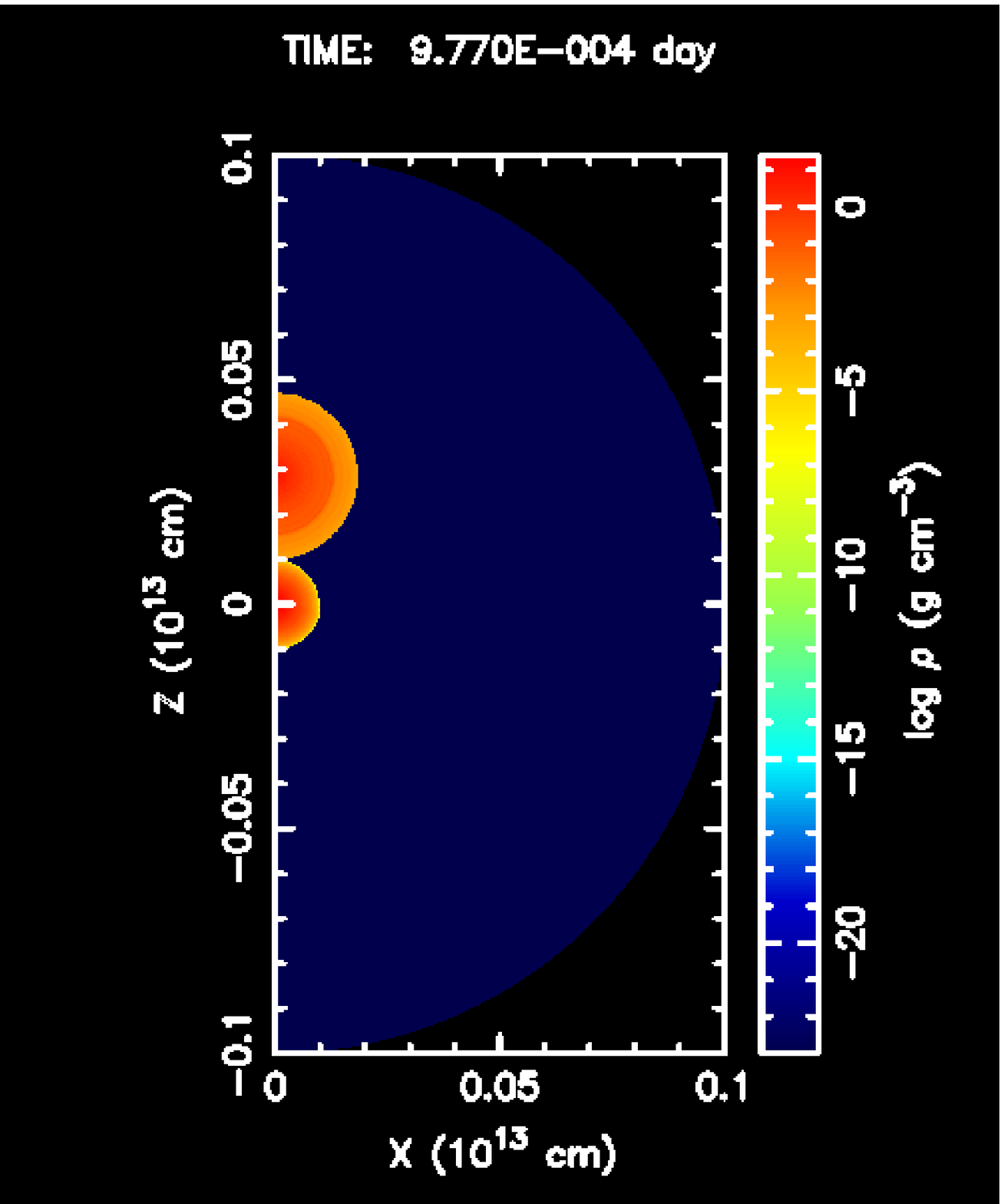}
  \includegraphics[width=0.3\linewidth]{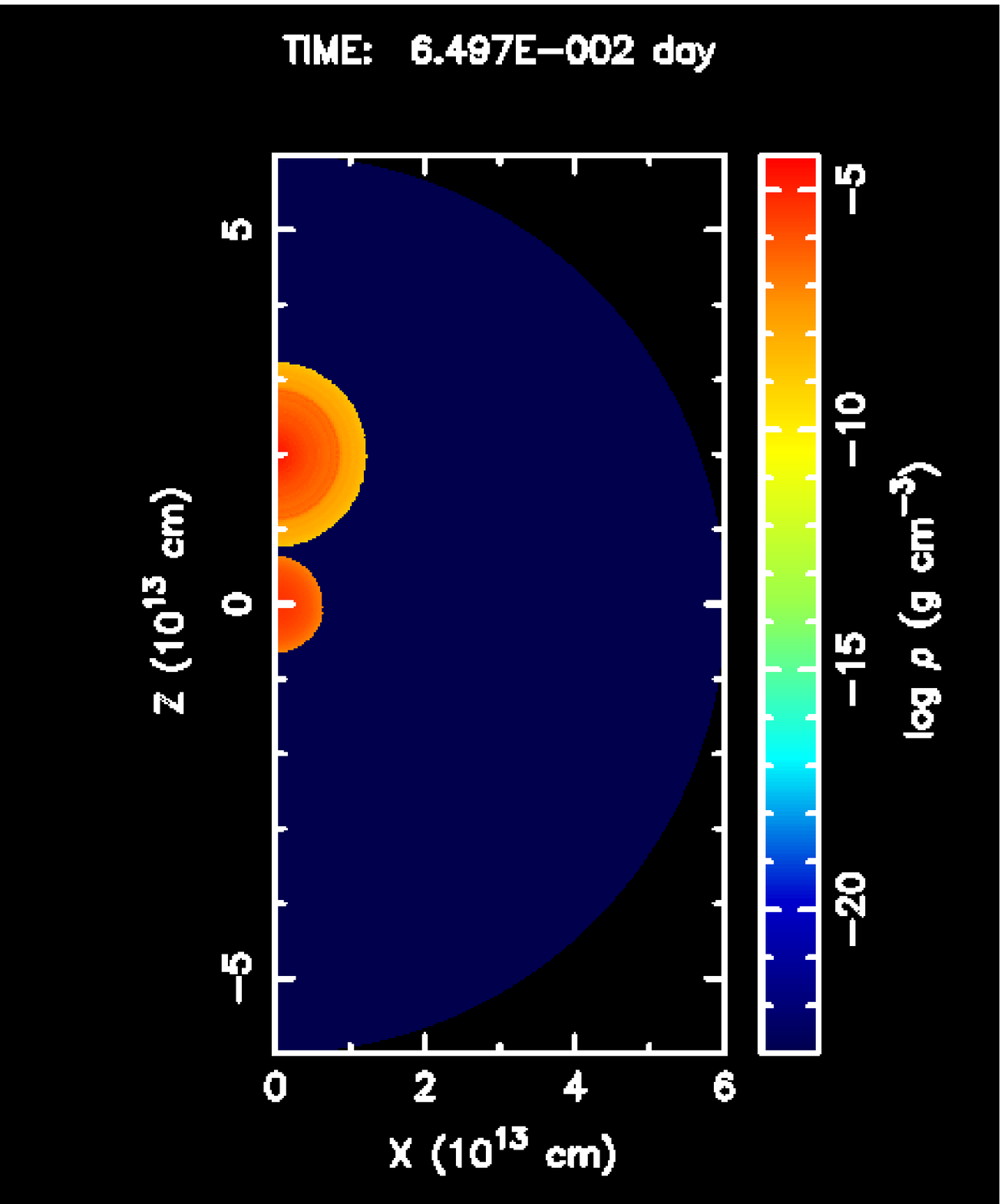}
  \includegraphics[width=0.3\linewidth]{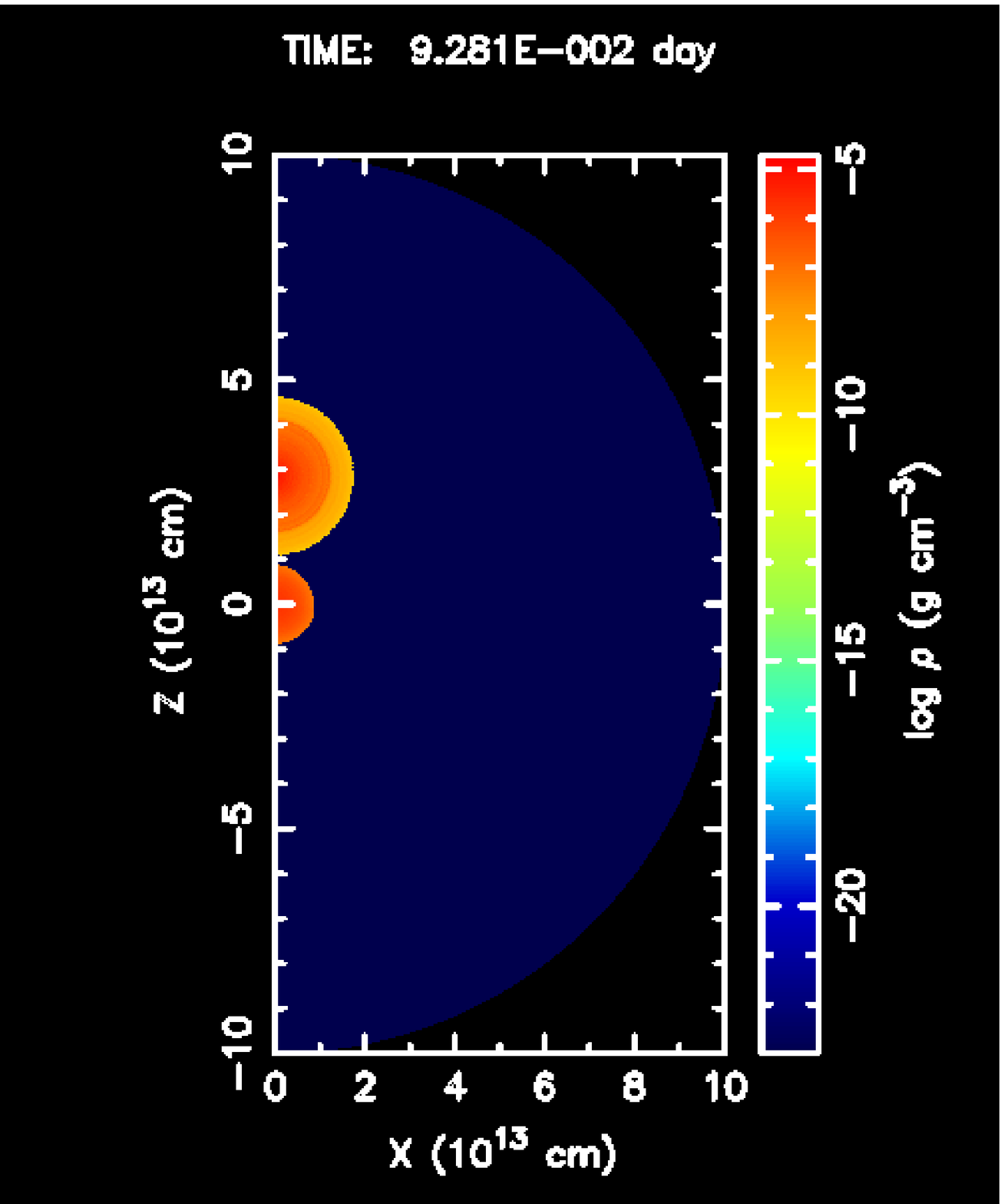}
  \end{center}
  \caption{Initial condition of each model. ---
  The left panel shows model MS.
  The center panel shows model RGa.
  The right panel shows model RGb.
  Note that scales are different in these models.
  }
  \label{study-initial2d.eps}
\end{figure*}

A companion star is located at the origin of the coordinate system.
We put the center of the expanding ejecta at a point $r=A$, $\theta = 0$.
The initial distributions of the density and  abundance of each element in the expanding ejecta are obtained from W7 model \citep{1984ApJ...286..644N}.
They are shown in Figure~\ref{study-w7.eps}.
Considered isotopes are
$\iso{12}{C}$,
$\iso{16}{O}$,
$\iso{20}{Ne}$,
$\iso{24}{Mg}$,
$\iso{28}{Si}$,
$\iso{32}{S}$,
$\iso{36}{Ar}$,
$\iso{40}{Ca}$,
$\iso{52}{Fe}$
and $\iso{56}{Ni}$.
Calculations are initiated at the moment when the ejecta touch the surface of the companion.
Figure~\ref{study-initial2d.eps} shows the initial density distribution of each model.

\subsection{Input physics}
Here,
we describe how to calculate gravity, transport coefficients, and nuclear reactions in equations~(\ref{hydro-mass})-(\ref{eqrhd-end}).
Since the gravity mainly affects the amount of the stripped material,
we assume that the companion star is the only source of gravity,
and further that
the gravitational field does not change with time.
About the transport process,
we take into account the free-free absorption and the Thomson scattering.
We use the Planck-mean of the free-free absorption coefficient as the bolometric absorption coefficient.
It is underestimated because we ignore absorption by bound-bound and bound-free transitions.

The ejecta are heated by gamma-ray from the radioactive decays of $\iso{56}{Ni}$ to $\iso{56}{Co}$, and $\iso{56}{Co}$ to $\iso{56}{Fe}$.
The life time of $\iso{56}{Ni}$ we use in our simulations is $\tau(\iso{56}{Ni}) = 8\unit{days}$ and the energy emitted from its decay $E(\iso{56}{Ni}) = 2.13\unit{MeV}$  \citep{1996snih.book.....A}.
For the $\iso{56}{Co}$ decay, we use $\tau(\iso{56}{Co}) = 111\unit{days}$ and $E(\iso{56}{Co}) = 4.56\unit{MeV}$.
The decay rates of isotopes are written as
\begin{eqnarray}
  D(\iso{56}{Ni}) &=& - m(\iso{56}{Ni})\,\frac{n(\iso{56}{Ni})}{\tau(\iso{56}{Ni})}
  ,\\
  D(\iso{56}{Co}) &=& m(\iso{56}{Ni})\,\frac{n(\iso{56}{Ni})}{\tau(\iso{56}{Ni})}
  - m(\iso{56}{Co})\,\frac{n(\iso{56}{Co})}{\tau(\iso{56}{Co})}
  ,\\
  D(\iso{56}{Fe}) &=& m(\iso{56}{Co})\,\frac{n(\iso{56}{Co})}{\tau(\iso{56}{Co})}
  ,
\end{eqnarray}
where $m(A)$ and $n(A)$
are the mass and number density of the isotope $A$, respectively.
Then, the gamma-ray energy generation rate $\mathcal{E}_\gamma$ is calculated by
\begin{equation}
  \mathcal{E}_\gamma =
  E(\iso{56}{Ni})\frac{n(\iso{56}{Ni})}{\tau(\iso{56}{Ni})}
  + E(\iso{56}{Co})\frac{n(\iso{56}{Co})}{\tau(\iso{56}{Co})}.
\end{equation}
We do not calculate the transport of emitted gamma-ray photons.
Instead, we assume that a fraction of the generated gamma-ray is immediately absorbed by the local material to the extent determined by the optical depth of the ejecta.
The heating rate $\epsilon$ is then written as
\begin{equation}
  \epsilon = \mathcal{E}_\gamma [1 - \exp(-\tau_\gamma)].
\end{equation}
Here $\tau_\gamma$ is the gamma-ray optical depth calculated by
\begin{equation}
  \tau_\gamma = \int \kappa_\gamma \rho dr,
\end{equation}
where $\kappa_\gamma$ is the gamma-ray absorption opacity.
Here the optical depth is integrated along the line passing the center of the ejecta.
\citet{1989ApJ...343..323F} studied the gamma-ray transport in SN Ia, assuming the gamma-ray is not scattered, only absorbed by the ejecta.
They obtain $\kappa_\gamma = 0.03\unit{cm^2g^{-1}}$ to reproduce the observed light curves of SNe Ia.
Although our assumption might be a crude approximation,
the total amount of thermal energy converted from the gamma-ray energy is almost correct
in the sense that it can reproduce the typical light curve of SN Ia and the result of \citet{2014ApJ...794...37M}.  

\subsection{Remapping}
Initially, the physical scale of the binary system is $10^{11}\unit{cm}$ in model MS, and $10^{13}\unit{cm}$ in models RGa and RGb.
On the other hand, at the maximum light ($\sim 20$ days after the explosion), the radius of the ejecta becomes $3 \times 10^{15} \unit{cm}$.
In order to trace the profiles of the ejecta and the stripped material,
we expand the calculation region according to the expansion of the ejecta.
When the size of the ejecta reaches 90\% of the size of the calculation region,
we increase the $r$-axis interval of computational grids by a factor of two.
The values on the outermost cells of the original region are substituted to fluid variables in newly created cells outside the original calculation region.

\section{Ray tracing}
As results of the above simulation,
we obtain snapshots of the spacial distribution of each variable.
Integrating the radiation transport equation along a specified line of sight on the snapshot,
we obtain light curves and spectra.
Here, we define the Cartesian coordinates $(x,y,z)$ of which the $z$-axis is along the line of sight.
Considering both thermal emission and isotropic scattering,
the surface brightness along the $z$-axis is written as
\begin{equation}
  \mathcal{I}_\nu(x,y) = \int (\alpha_\nu+\sigma_\nu) (S_\nu-I_\nu)\,dz,
\end{equation}
where
$I_\nu$ is the intensity along the $z$-axis.
$S_\nu$ is the source function defined by
\begin{equation}
  S_\nu = \frac{\alpha_\nu B_\nu(T) + \sigma_\nu J_\nu}{\alpha_\nu + \sigma_\nu},
\end{equation}
where
$B_\nu(T)$ is the Planck function, 
and $J_\nu$ is the mean intensity.
Then, the isotropic luminosity is calculated by
\begin{equation}
  L^\mathrm{iso}_\nu = 4\pi \int \int \mathcal{I}_\nu(x,y)\,dx\,dy.
\end{equation}

From the radiation hydrodynamic simulation,
we have distributions of the bolometric mean intensity $J = (c/4\pi)E$.
Then, the bolometric source function $S$ is calculated by
\begin{equation}
  S = \frac{\alpha c a T^4/4\pi + \sigma J}{\alpha + \sigma}.
\end{equation}
Here, $\alpha$ and $\sigma$ the bolometric absorption and scattering coefficients, which are equal to those used in the above simulation.
Computing the bolometric luminosity from each snapshot, we obtain the bolometric light curve.

To obtain a spectrum, we need to calculate the monochromatic source function $S_\nu$.
Though we need the monochromatic mean intensity $J_\nu$ to calculate $S_\nu$,
we know only the value of bolometric mean intensity $J$ from the simulation.
To evaluate $J_\nu$, we assume that $J_\nu$ takes a form of the Planck function,
\begin{equation}
  J_\nu = B_\nu(T_\mathrm{rad}),\quad J = c a T_\mathrm{rad}^4 / 4\pi.
\end{equation}
To determine the absorption and scattering coefficients,
we take into account
Thomson scattering,
free-fee absorption,
and bound-bound absorption by relevant atomic lines.
Ionization states of each isotope are calculated by the Saha equation.
To calculate the optical depth of the line absorption in moving media, we apply the Sobolev approximation \citep{1978ApJ...219..654R}.
In this study, we only concern with hydrogen features in optical spectra.
Then, we calculate only H$\alpha$, C II, and Si II lines.
These lines are located in a narrow range of the rest wavelengths between 6300\,\AA\ and 6600\,\AA.
Table~\ref{lines.tbl} shows spectral lines considered in this study.

\begin{table}
  \caption{Parameters of spectral lines.
  ---
  $f$ is the oscillator strength.
  $g_i$ and $E_i$ are the statistical weight and the level energy of the lower level $i$.
  $g_j$ and $E_j$ are those of the upper level $j$.
  }
  \label{lines.tbl}
  \begin{center}
    \begin{tabular}{lrrrrr}
      \hline
      \hline
      Lines & $\log(gf)$ & $E_i$ (eV) & $E_j$ (eV) & $g_i$ & $g_j$
      \\
      \hline
      H I $\lambda$6562.81\AA & 0.710 & 10.198 & 12.087 & 8 & 18
      \\
      C II $\lambda$6578.05\AA & -0.021 & 14.449 & 16.333 & 2 & 4
      \\
      C II $\lambda$6582.88\AA & -0.323 & 14.449 & 16.332 & 2 & 2
      \\
      Si II $\lambda$6347.10\AA & 0.149 & 8.121 & 10.073 & 2 & 4
      \\
      Si II $\lambda$6371.36\AA & -0.082 & 8.121 & 10.066 & 2 & 2
      \\
      \hline
    \end{tabular}
  \end{center}
\end{table}

When neutral hydrogen is located outside the photosphere where the temperature is lower than the photospheric temperature,
H$\alpha$ line might appear as absorption in the spectrum.

\section{Results}
Results from the simulation and ray-tracing explained in the previous sections are presented in the following subsections.

\subsection{Stripped mass}
From the above simulation, we obtain spatial structures of the ejecta and the stripped material.
The stripped mass from the companion star is calculated by adding up the mass of hydrogen rich gas with the fluid velocity exceeding the escape velocity.
The masses are $0.45M_\odot$ in model MS, $0.51M_\odot$ in model RGa,  and $0.49M_\odot$ in model RGb.
In model MS,
the amount is overestimated, compared to the previous studies (about $0.01-0.02M_\odot$ in \citet{2008AA...489..943P}).
This is because we do not have enough resolution in the low velocity region around the companion star due to the remapping process.
On the other hand, in models RGa and RGb,
the amount is slightly smaller than the previous studies \citep{1975ApJ...200..145W,2000ApJS..128..615M}.
Nevertheless, the trend that the entire envelope is stripped by the impact
is consistent with these studies.

\subsection{Structures obtained from the simulation}
\begin{figure*}
   \includegraphics[width=0.5\textwidth]{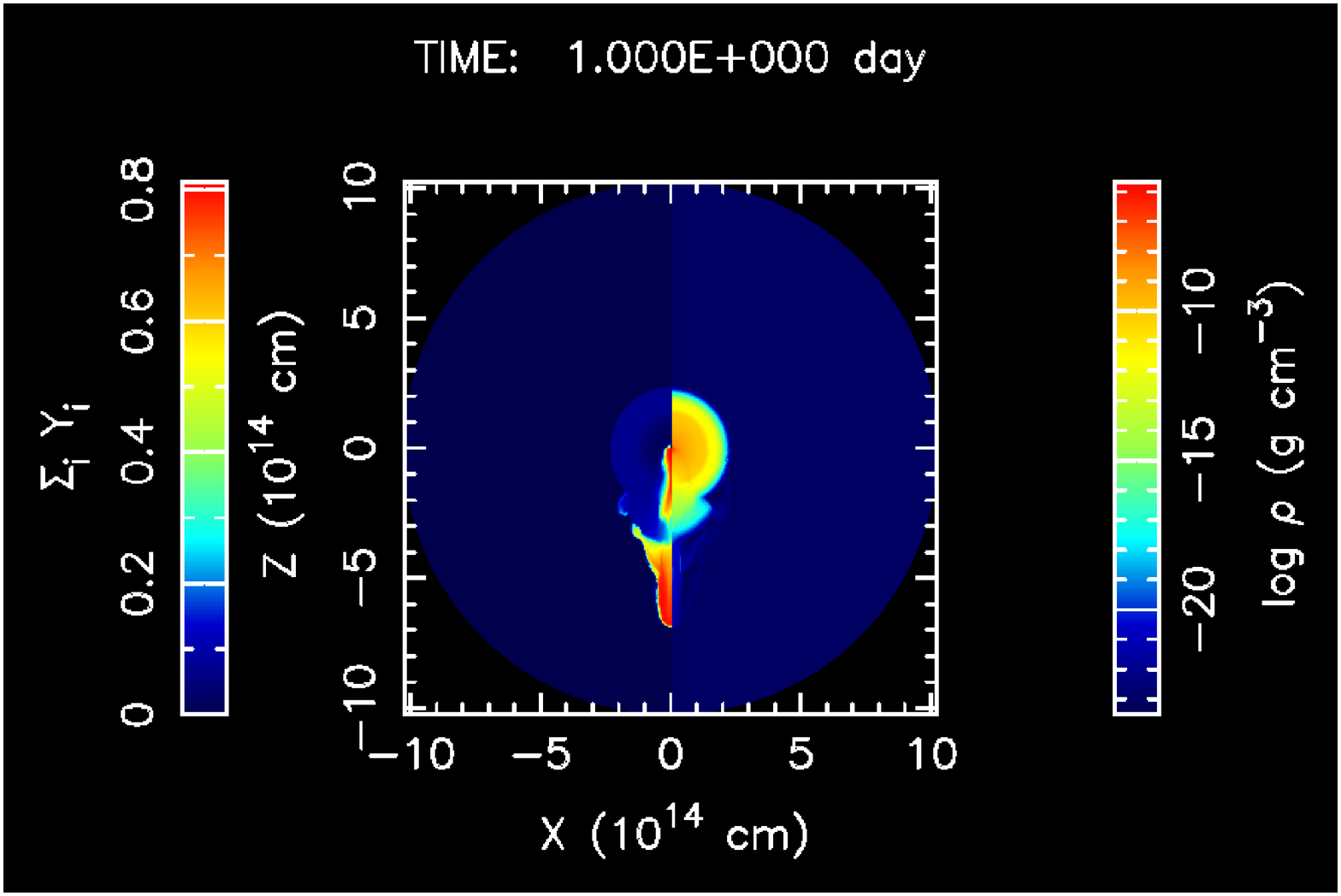}
   \includegraphics[width=0.5\textwidth]{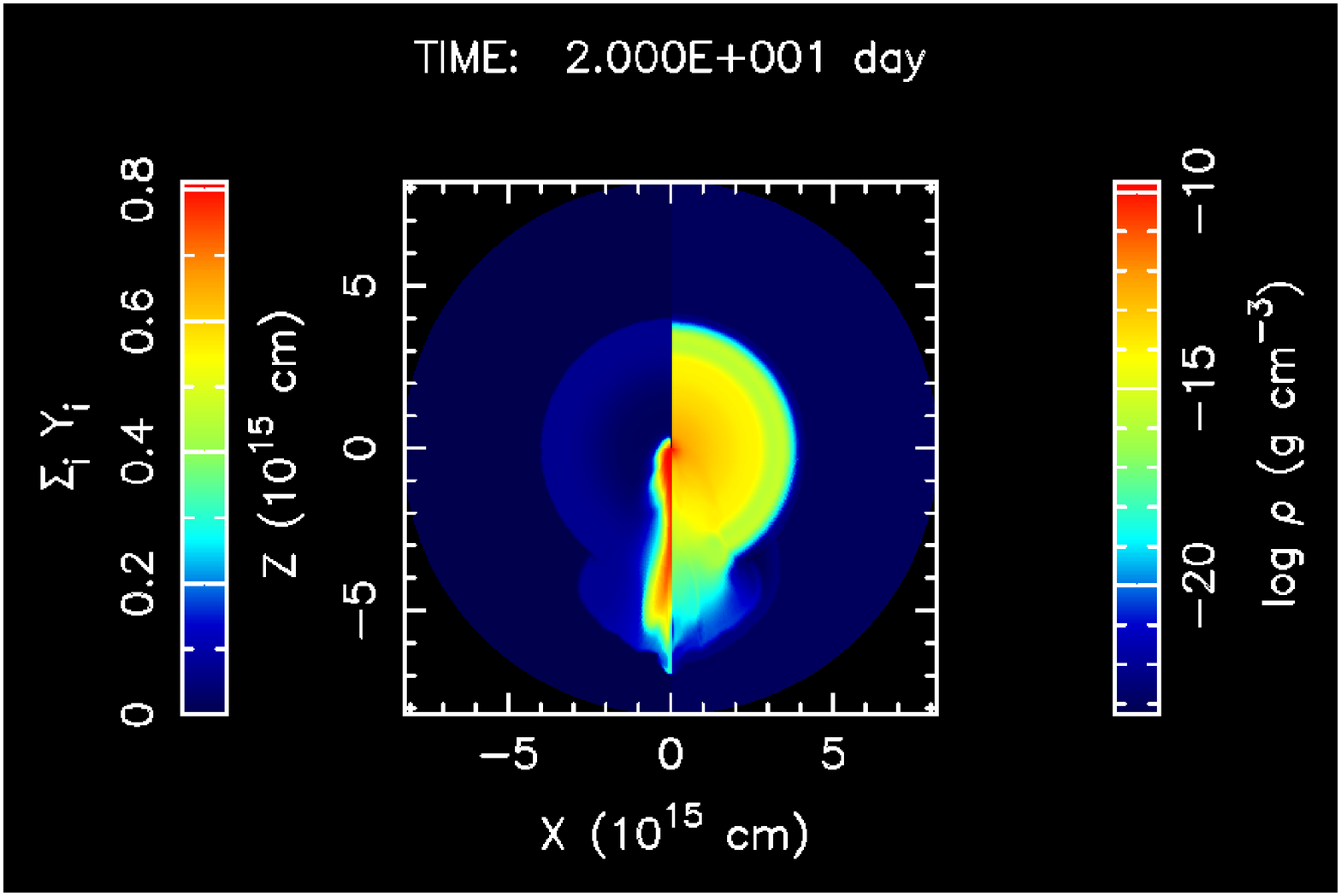}
  \caption{Density and abundance structures in model MS at days 1 and 20 after the explosion. ---
  In each panel, the number abundance is shown on the left side, and the density is on the right side.
  The red region on the left side is rich in hydrogen.
  }
  \label{pict2d-MS.eps}
\end{figure*}

\begin{figure*}
  \includegraphics[width=0.5\textwidth]{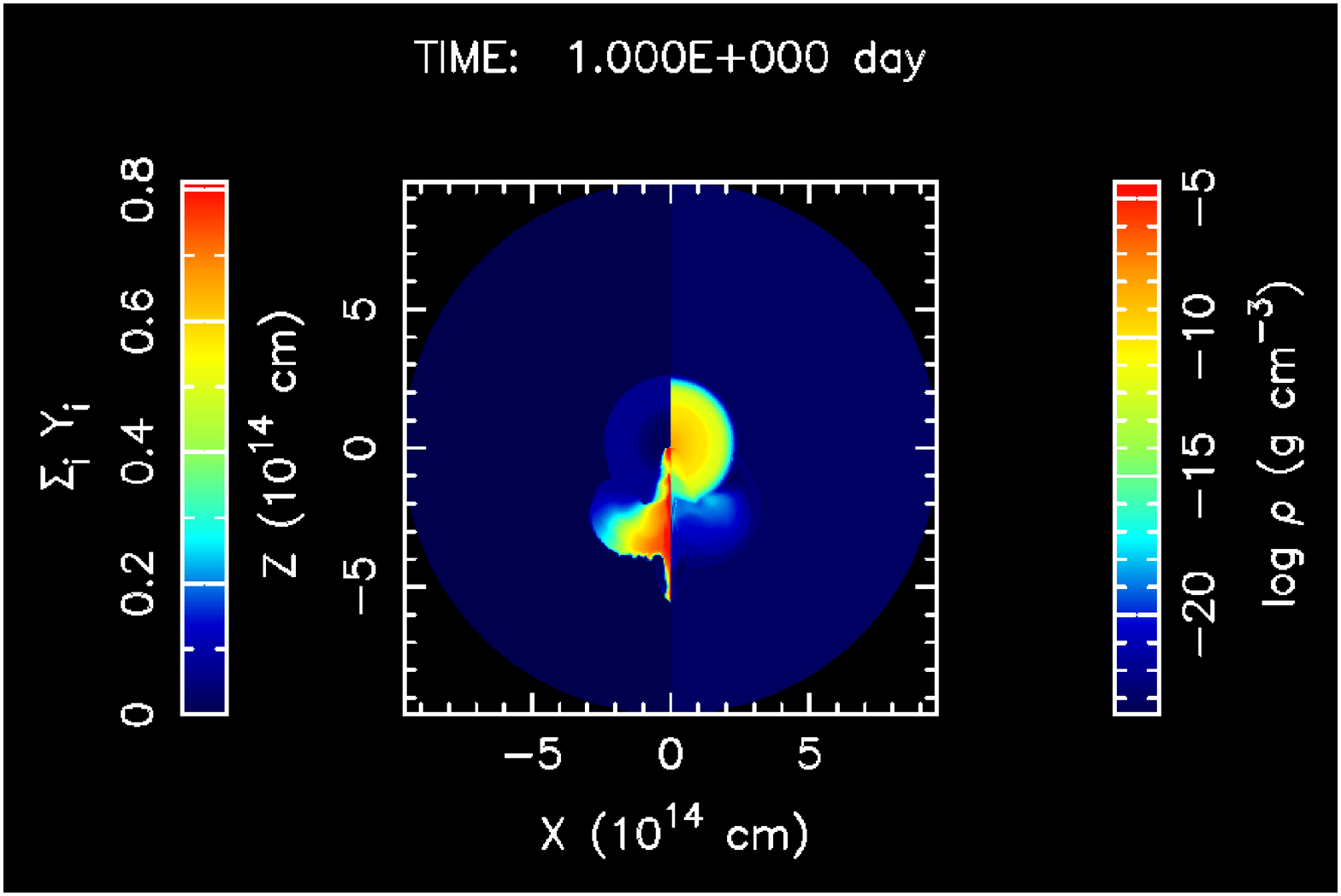}
  \includegraphics[width=0.5\textwidth] {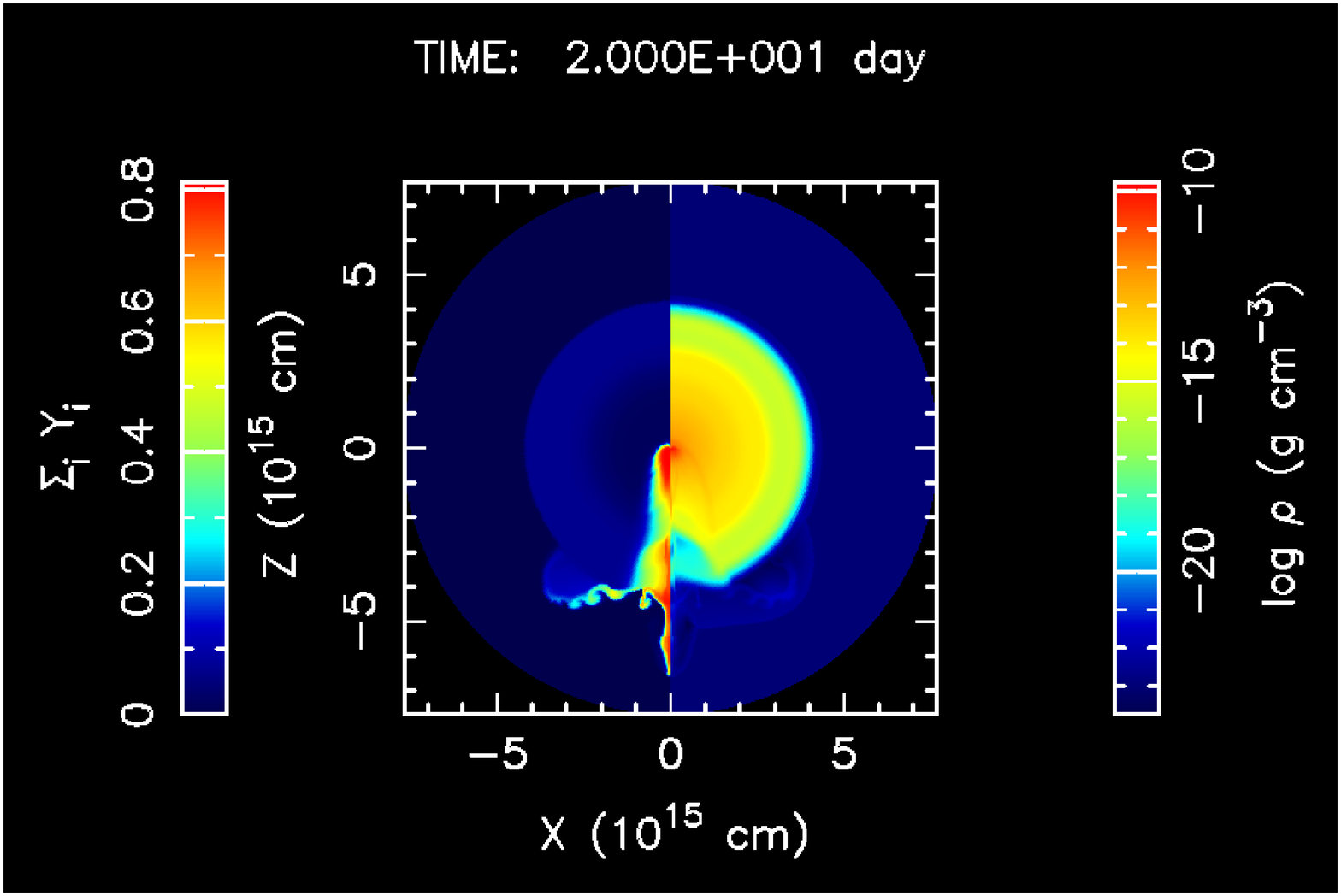}
  \caption{Density and abundance structures in model RGa at days 1 and 20 after the explosion. ---
  In each panel, the number abundance is shown on the left side, and the density is no the right side.
  The red region on the left side is rich in hydrogen.
  }
  \label{pict2d-RGa.eps}
\end{figure*}

\begin{figure*}
  \includegraphics[width=0.5\textwidth]{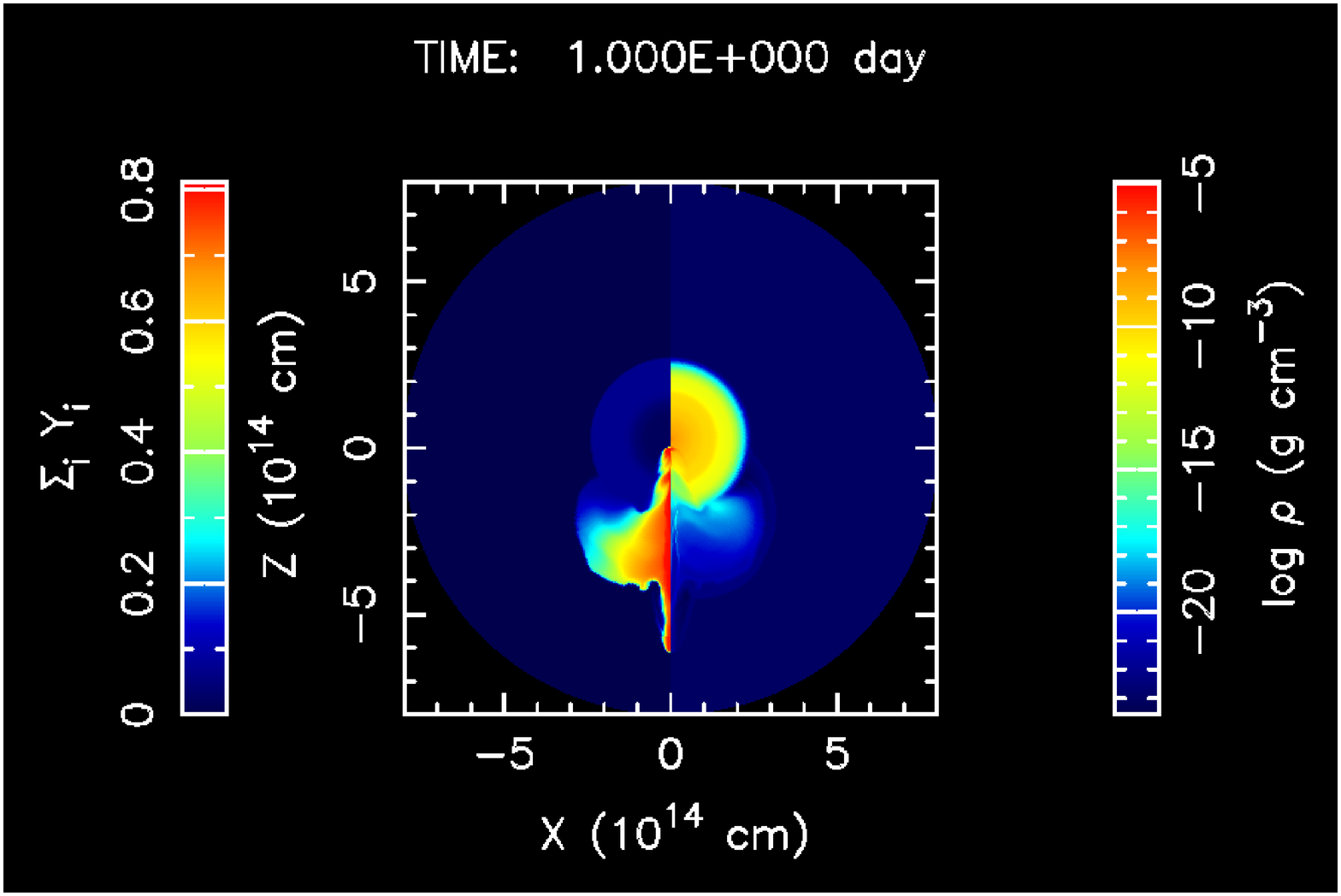}
  \includegraphics[width=0.5\textwidth]{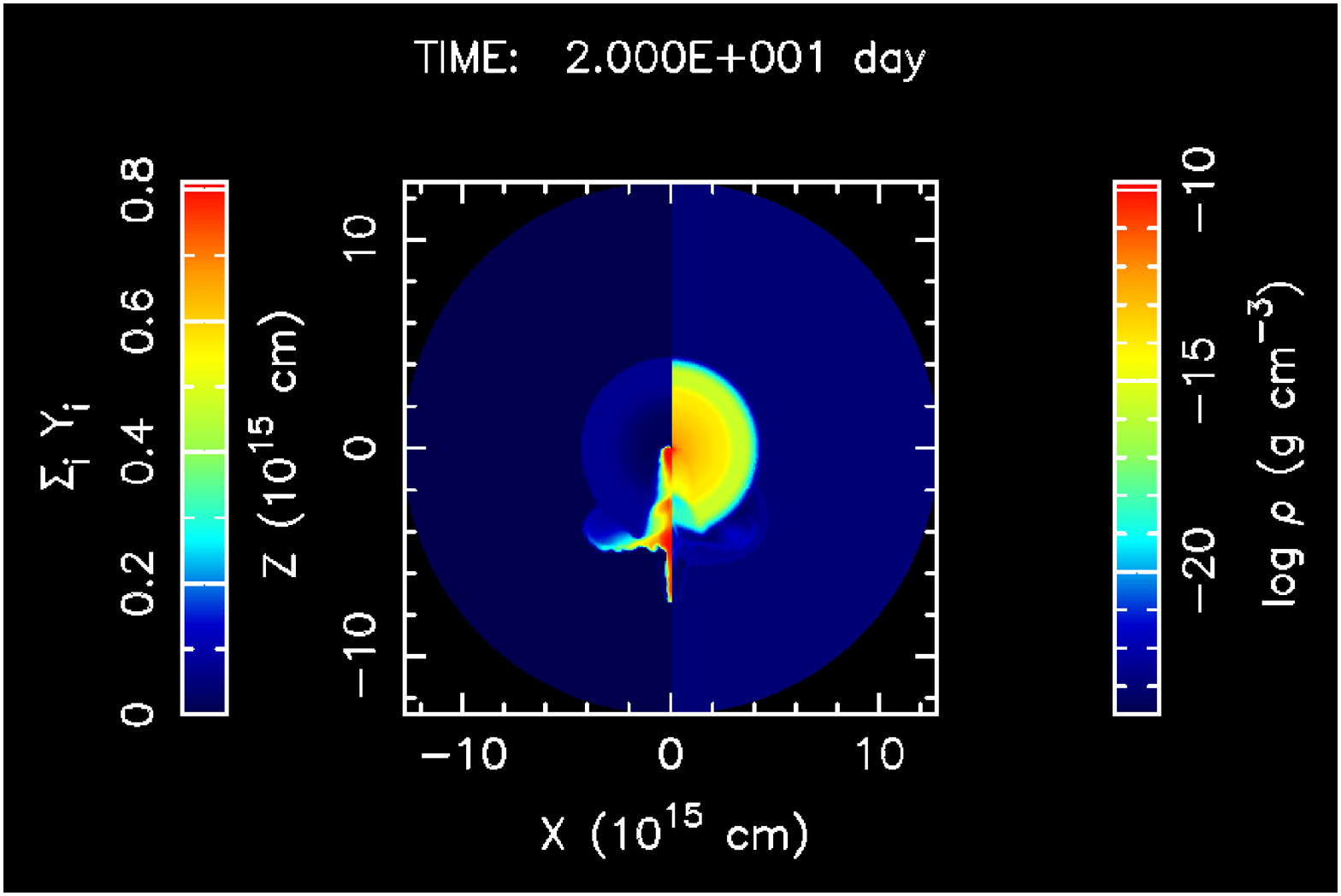}
  \caption{Density and abundance structures in model RGb at days 1 and 20 after the explosion. ---
  In each panel, the number abundance is shown on the left side, and the density is on the right side.
  The red region on the left side is rich in hydrogen.
  }
  \label{pict2d-RGb.eps}
\end{figure*}

Figures~\ref{pict2d-MS.eps}-\ref{pict2d-RGb.eps} show distributions of density $\rho$ and total number abundance $\sum_i(X_i/A_i)$ at days 1 and 20 after the explosion in each model.
As shown in the figures,
the hydrogen-rich material is stripped along the line $\theta = \pi$.
At day 20,
the ejecta and the stripped material are already expanding homologously.
A cavity is created around the axis $\theta = \pi$
with the spread angle of $\sim\pi/6\unit{rad}$.
%The angle range that the cavity spreads is about $\pi/6\unit{rad}$.
If we view this SN with a viewing angle in this range,
the stripped hydrogen is not hidden by the surrounding ejecta.

\begin{figure*}
  \begin{center}
    \includegraphics[width=0.5\textwidth]{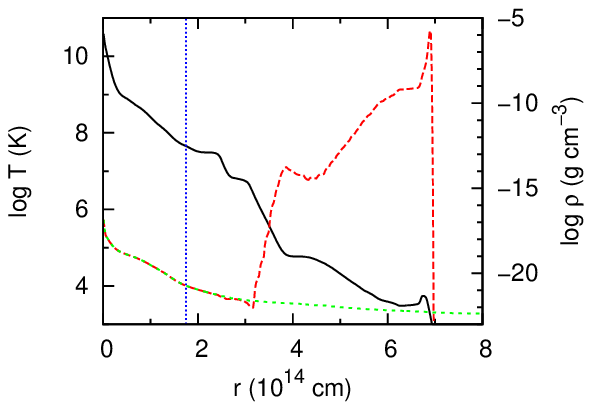}
   \includegraphics[width=0.5\textwidth]{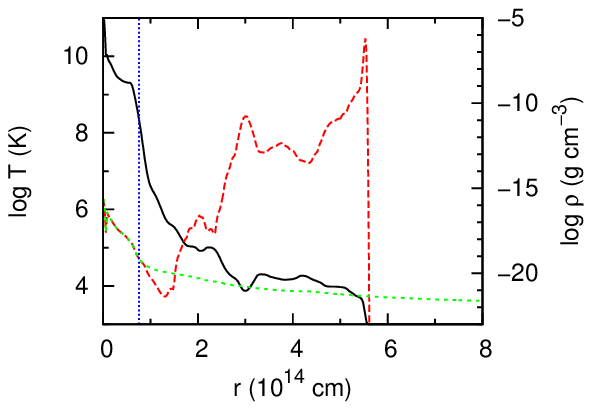}
   \includegraphics[width=0.5\textwidth]{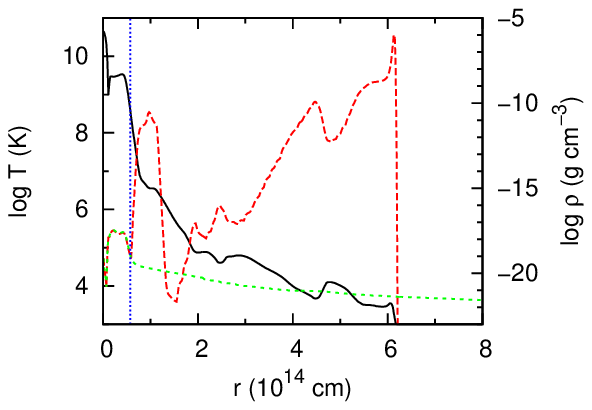}
  \end{center}
  \caption{Density and temperature structures along $\theta = \pi$ at day 1 after the explosion in each model. ---
  The top panel shows model MS.
  The center panel shows model RGa.
  The bottom panel shows model RGb.
    The solid line is density, long-dashed line is gas temperature, and short-dashed line is radiation temperature, respectively.
  The dotted line shows position of the photosphere.
  }
  \label{pict1d.eps}
\end{figure*}

Figure~\ref{pict1d.eps} shows one-dimensional structures of the density $\rho$, the gas temperature $T$, and the radiation temperature $T_\mathrm{rad}$ along the line $\theta = \pi$ at day 1.
We calculate the position of the photosphere where thermalization length is equal to unity,
which is indicated by the dotted vertical line.
The photospheric temperature is equal to
$1.0 \times 10^4\unit{K}$ for model MS,
$5.2 \times 10^4\unit{K}$ for model RGa,
and $6.2 \times 10^4\unit{K}$ for model RGb.
\citet{2010ApJ...708.1025K} analytically estimated the effective temperature of the prompt emission as
\begin{equation}
  T_\mathrm{eff} = 2.5\times10^4 a_{13}^{1/4}
  \kappa_{e}^{-35/36} t_\mathrm{day}^{-37/72}\unit{K},
\end{equation}
where $a_{13} = A/10^{13}\unit{cm}$, $\kappa_{e}$ is the electron scattering opacity,
and $t_\mathrm{day}$ is the time in days.
At $t=1\unit{day}$,
$T_\mathrm{eff}$ is estimated as $5.0\times10^{4}\unit{K}$ for model MS, 
$1.4\times10^{5}\unit{K}$ for model RGa,
and $1.6\times10^{5}\unit{K}$ for model RGb,
In comparison, the temperatures become lower in our simulation.
This is because the collision heats the gas first in our simulations, which separately treat the temporal evolutions of the radiation and gas energy densities. The subsequent emission from the gas is not enough to raise the photospheric temperature as high as this formula due to the low densities.

\subsection{Light curves}
\begin{figure*}
  \begin{center}
   \includegraphics[width=0.5\textwidth]{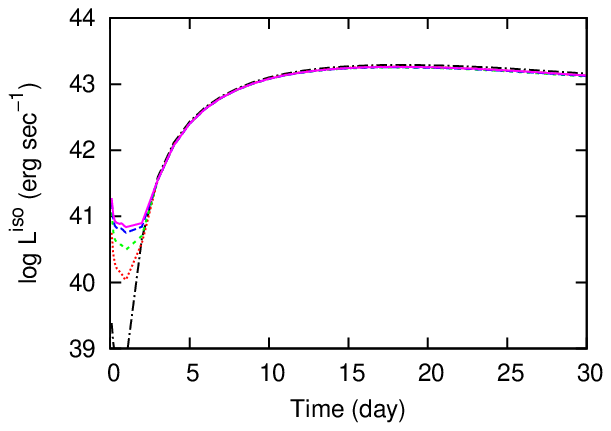}
   \includegraphics[width=0.5\textwidth]{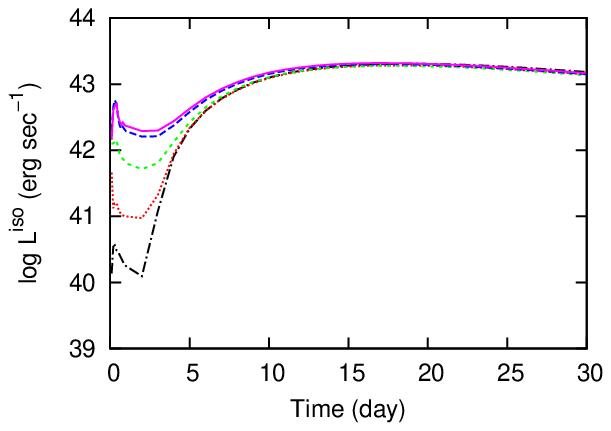}
   \includegraphics[width=0.5\textwidth]{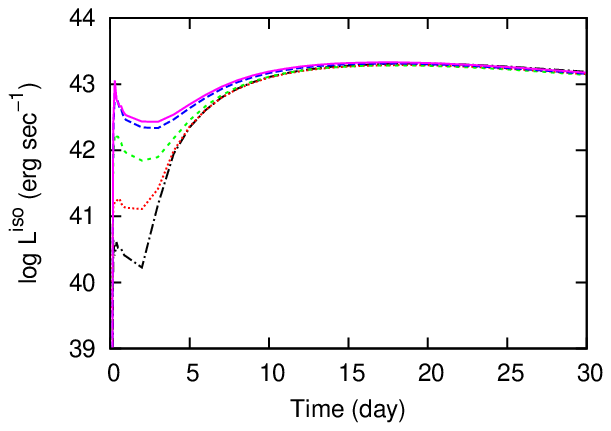}
  \end{center}
  \caption{Bolometric light curves with different viewing angles. ---
  The top panel shows model MS.
  The center panel shows model RGa.
  The bottom panel shows model RGb.
  The dash-dotted, dotted, short-dashed, long-dashed, and solid lines show light curves when we observe the SN with the viewing angle $\theta = 0$, $\pi/2$, $2\pi/3$, $5\pi/6$, and $\pi$, respectively.
  }
  \label{lcurve.eps}
\end{figure*}

\begin{figure*}
  \begin{center}
\includegraphics[width=0.5\textwidth]{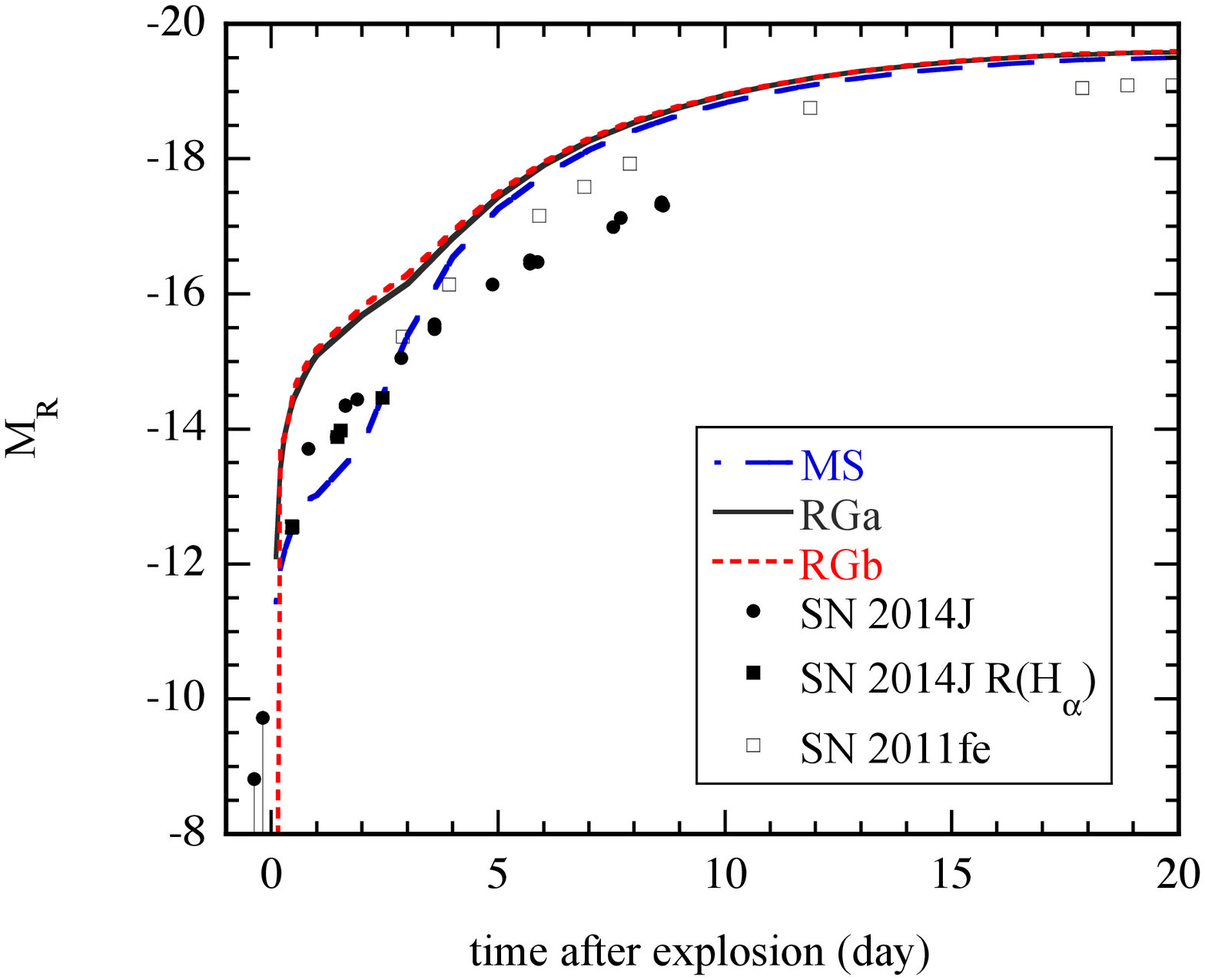}
 \includegraphics[width=0.5\textwidth]{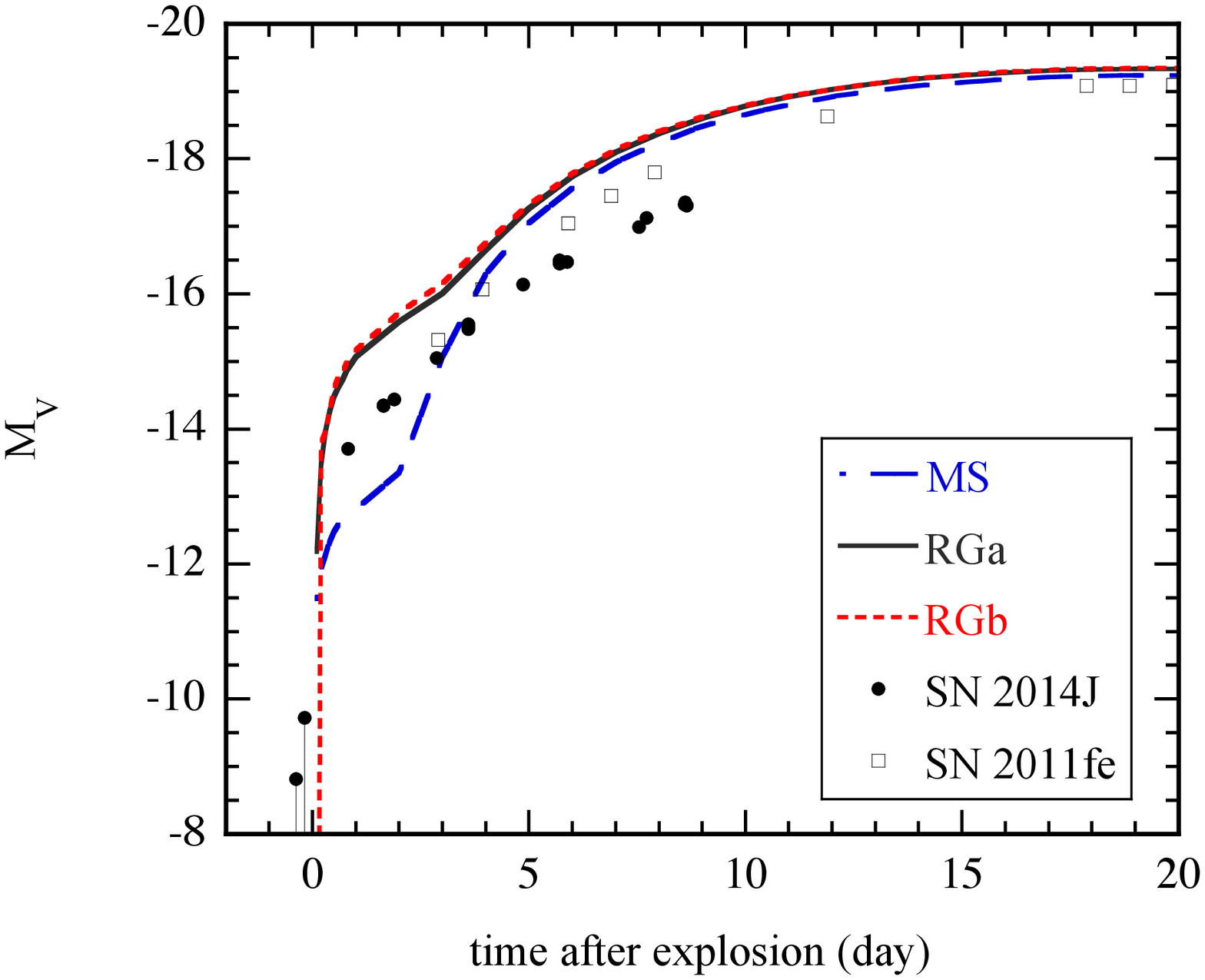}
  \end{center}
  \caption{Model light curves in the R band (top panel) and V band (bottom panel) viewing from the side of the companion stars.   Light curves of well observed type Ia supernovae 2011fe \citep{2012JAVSO..40..872R} and 2014J \citep{2014ApJ...783L..24Z} are also plotted for comparison. Note that the points of SN 2014J are of unfiltered observations except for filled squares in the top panel. The filled squares denote the $R$ magnitudes estimated from detections using the narrowband H$_\alpha$ filters. The distance to SN 2014J is assumed to be 3.5 Mpc.}
  \label{rvcurve.eps}
\end{figure*}

Bolometric light curves with viewing angles $\theta = 0$, $\pi/2$, $2\pi/3$, $5\pi/6$ and $\pi$ are shown in Figure~\ref{lcurve.eps}.
The collision radiates prompt emission especially in the range of $\theta > 5\pi/6$.
\citet{2010ApJ...708.1025K} analytically estimated bolometric luminosities
due to the collision
at day 1
to be
$3\times 10^{41}\unit{erg~s^{-1}}$ for the $1M_\odot$ main sequence model,
$2\times 10^{42}\unit{erg~s^{-1}}$ for the $6M_\odot$ main sequence model,
and $2\times 10^{43}\unit{erg~s^{-1}}$ for the $1M_\odot$ red giant model.
In comparison with our results, the luminosity is equal to
$6.9\times 10^{40}\unit{erg~s^{-1}}$ for model MS,
$2.3\times 10^{42}\unit{erg~s^{-1}}$ for model RGa,
and $3.4\times 10^{42}\unit{erg~s^{-1}}$ for model RGb.
The luminosities in models MS, RGa are dimmer than the previous study.
This is because the temperature of emission region becomes lower than their results.
In the analysis of \citet{2010ApJ...722.1691H},
the luminosity in the $6M_\odot$ main sequence model
is the threshold above which a progenitor system is ruled out from the primary source of SNe Ia.
The luminosity in model RGa is on this threshold.
A model with a slightly longer separation as model RGb
is to be ruled out because the emission can yet be detected.
Here one should note that our models may underestimate the strength of the coupling between the gas and radiation because the free-free emission and the energy exchange by Compton scattering, which our models do not include, accelerate the equilibration as pointed out by \citet{1976ApJS...32..233W, 2010ApJ...716..781K}. As a result, we may significantly underestimate the luminosity due to the shock heating. More elaborate numerical studies including the Comptonization are needed to assess this effect.
In Figure~\ref{rvcurve.eps}, we compare our results of model light curves in the R and V bands viewing from the side of the companion star ($\theta=\pi$) with two well observed type Ia SNe  2011fe and 2014J. Though the early light curve of SN 2014J is of unfiltered observations or of approximate $R$ magnitudes estimated from narrowband detections using H$_\alpha$ filters\citep{2014ApJ...784L..12G, 2014arXiv1410.1363G}, the shape indicates some influence from the collision between the SN ejecta and the companion star. Therefore SN 2014J might have a red giant as the ex-companion star and prefers the SD scenario. Note that we do not intend to fit the observed light curve of SN 2014J with our current models. A detailed comparison will be discussed elsewhere.

\subsection{Spectra}%need discussion in relation to Maeda+2014
\begin{figure*}
  \begin{minipage}{0.5\linewidth}
    \includegraphics[width=0.9\textwidth]{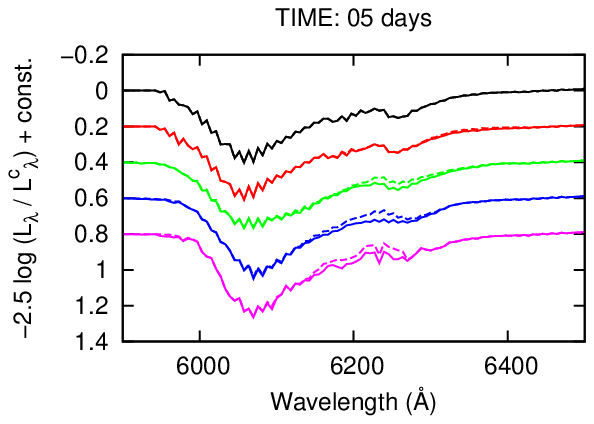}
  \end{minipage}
  \begin{minipage}{0.5\linewidth}
   \includegraphics[width=0.9\textwidth]{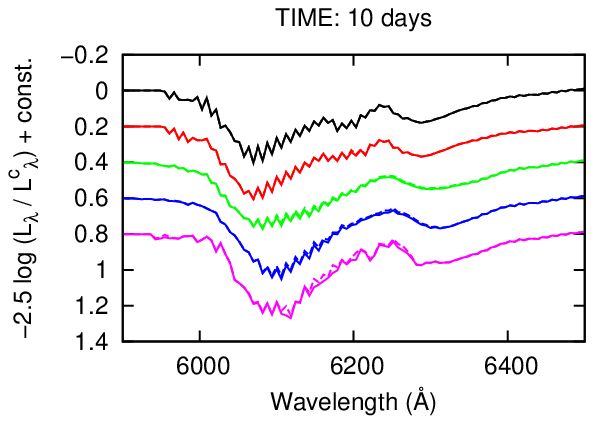}
  \end{minipage}
  \begin{minipage}{0.5\linewidth}
   \includegraphics[width=0.9\textwidth]{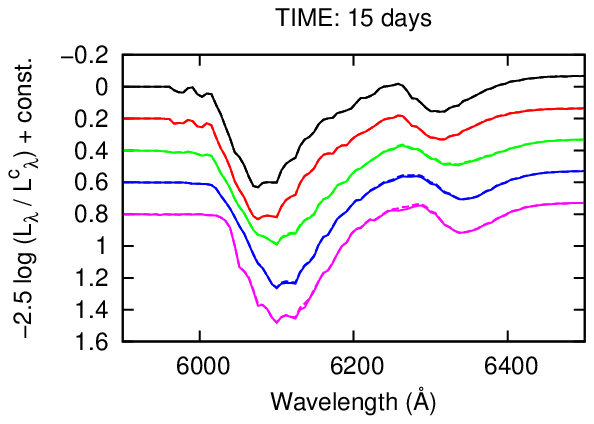}
  \end{minipage}
  \begin{minipage}{0.5\linewidth}
   \includegraphics[width=0.9\textwidth]{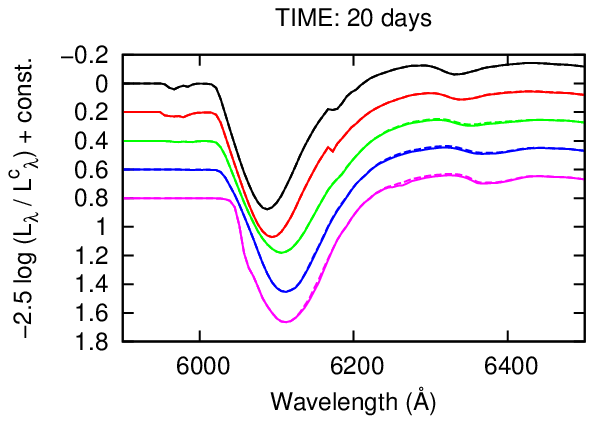}
  \end{minipage}
  \begin{minipage}{0.5\linewidth}
   \includegraphics[width=0.9\textwidth]{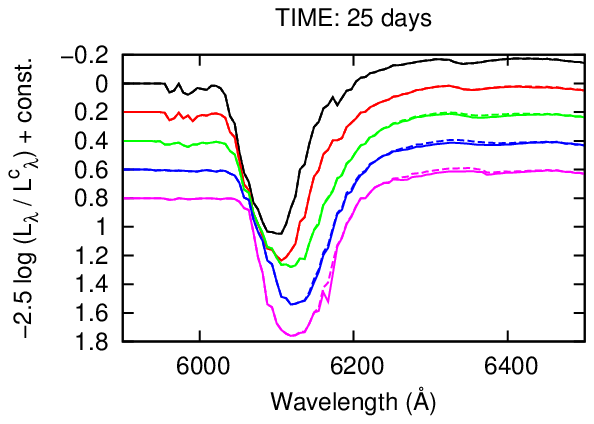}
  \end{minipage}
  \begin{minipage}{0.5\linewidth}
   \includegraphics[width=0.9\textwidth]{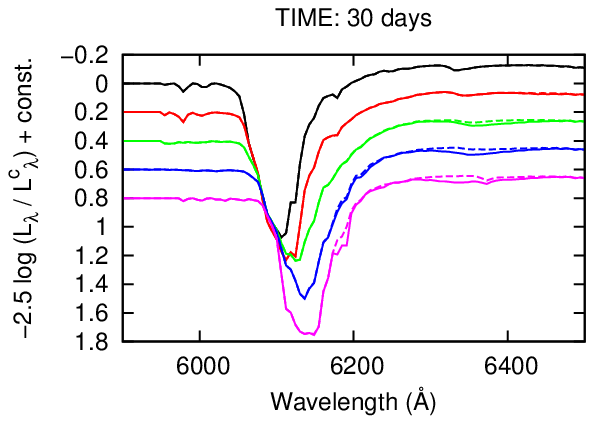}
  \end{minipage}
  \begin{minipage}{0.5\linewidth}
   \includegraphics[width=0.9\textwidth]{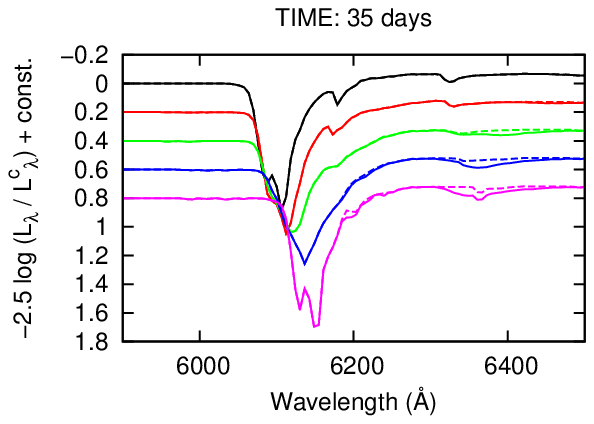}
  \end{minipage}
  \begin{minipage}{0.5\linewidth}
   \includegraphics[width=0.9\textwidth]{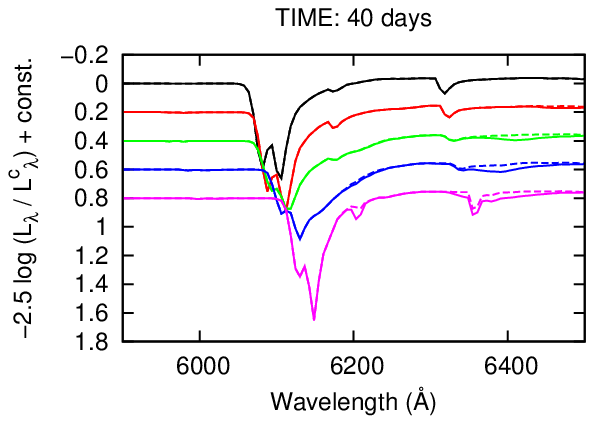}
  \end{minipage}
  \caption{Spectra  normalized by the continuum for model MS
  at days 5, 10, 15, 20, 25, 30, 35 and 40
  after the explosion.
  ---
  The black, red, green, blue, and magenta lines show spectra when we observe the SN with viewing angles $\theta = 0$, $2\pi/3$, $5\pi/6$, $11\pi/12$, and $\pi$, respectively.
  The solid lines are calculated including all spectral lines mentioned in the text, while the dashed lines without H$\alpha$ lines.
  }
  \label{spectra-MS-2.eps}
\end{figure*}

\begin{figure*}
  \begin{minipage}{0.5\linewidth}
   \includegraphics[bb=0 0 180 126, width=0.9\textwidth]{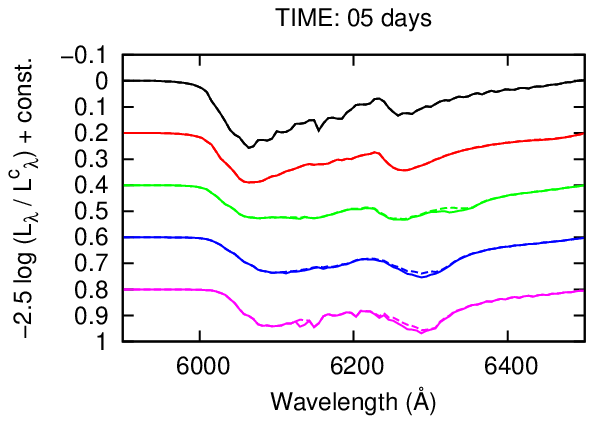}
  \end{minipage}
  \begin{minipage}{0.5\linewidth}
   \includegraphics[bb=0 0 180 126, width=0.9\textwidth]{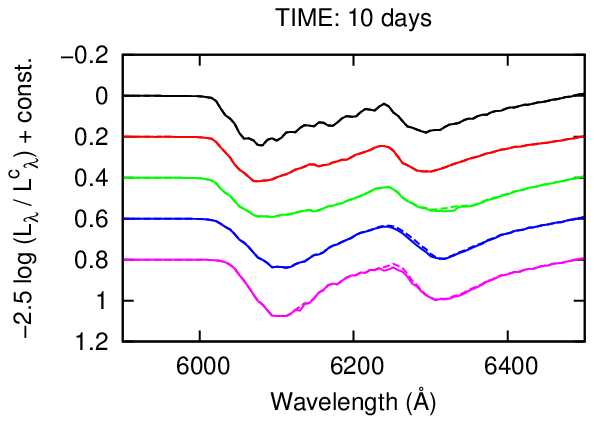}
  \end{minipage}
  \begin{minipage}{0.5\linewidth}
   \includegraphics[bb=0 0 180 126, width=0.9\textwidth]{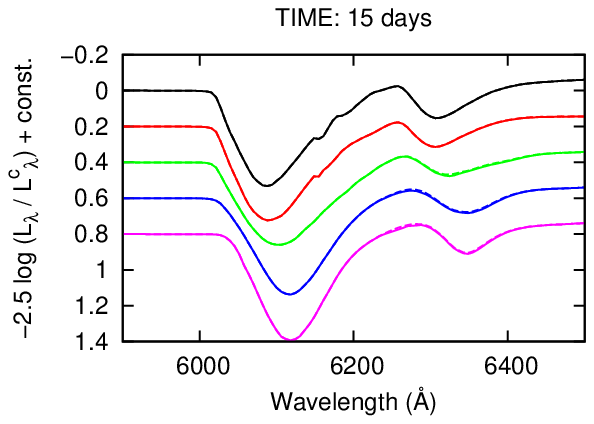}
  \end{minipage}
  \begin{minipage}{0.5\linewidth}
   \includegraphics[bb=0 0 180 126, width=0.9\textwidth]{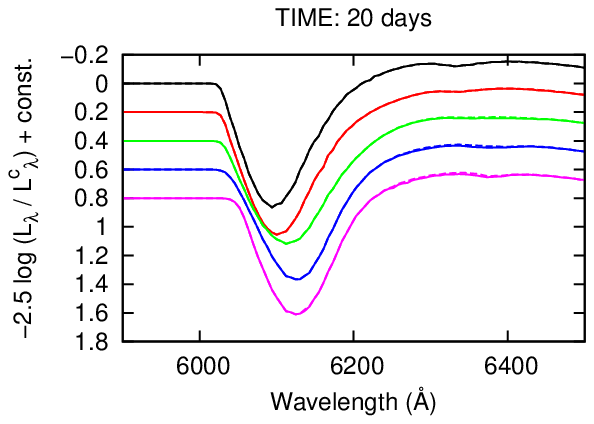}
  \end{minipage}
  \begin{minipage}{0.5\linewidth}
   \includegraphics[bb=0 0 180 126, width=0.9\textwidth]{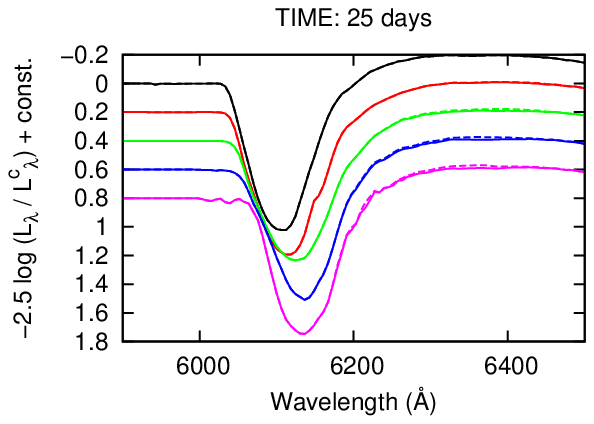}
  \end{minipage}
  \begin{minipage}{0.5\linewidth}
   \includegraphics[bb=0 0 180 126, width=0.9\textwidth]{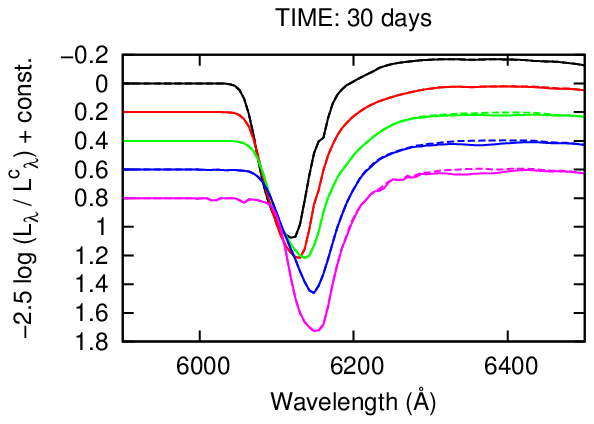}
  \end{minipage}
  \begin{minipage}{0.5\linewidth}
   \includegraphics[bb=0 0 180 126, width=0.9\textwidth]{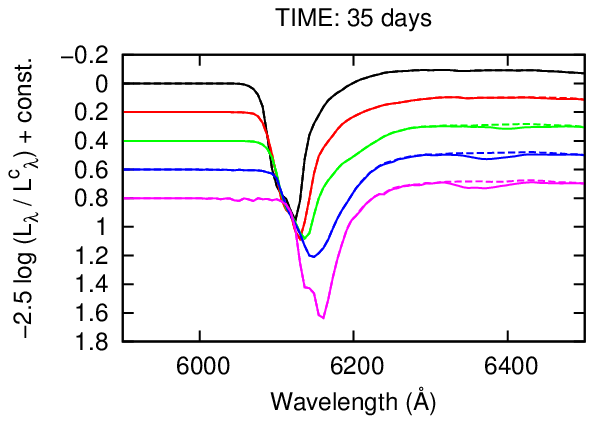}
  \end{minipage}
  \begin{minipage}{0.5\linewidth}
   \includegraphics[bb=0 0 180 126, width=0.9\textwidth]{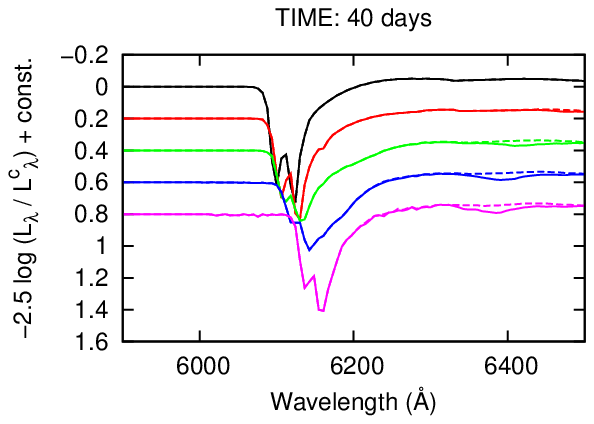}
  \end{minipage}
  \caption{Spectra in model RGa
  at day 5, 10, 15, 20, 25, 30, 35 and 40
  after the explosion.
  ---
  The black, red, green, blue, and magenta lines show light curves when we observe the SN with the viewing angle $\theta = 0$, $2\pi/3$, $5\pi/6$, $11\pi/12$, and $\pi$, respectively.
  The solid lines are calculated including all spectral lines.
  Besides, the dashed lines are calculated without H$\alpha$ lines.
  They are normalized by the continuum spectra.
  }
  \label{spectra-RG-2.eps}
\end{figure*}

\begin{figure*}
  \begin{minipage}{0.5\linewidth}
   \includegraphics[width=0.9\textwidth]{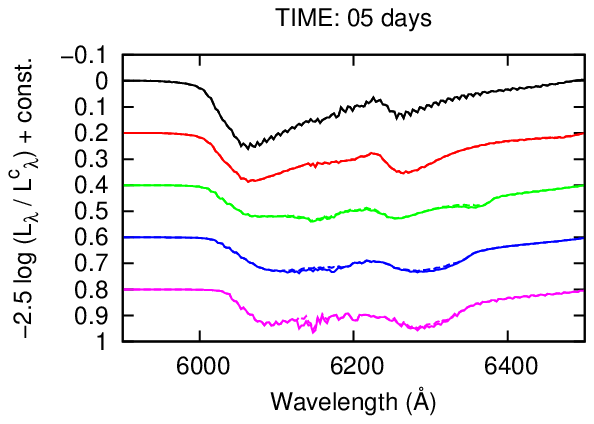}
  \end{minipage}
  \begin{minipage}{0.5\linewidth}
   \includegraphics[width=0.9\textwidth]{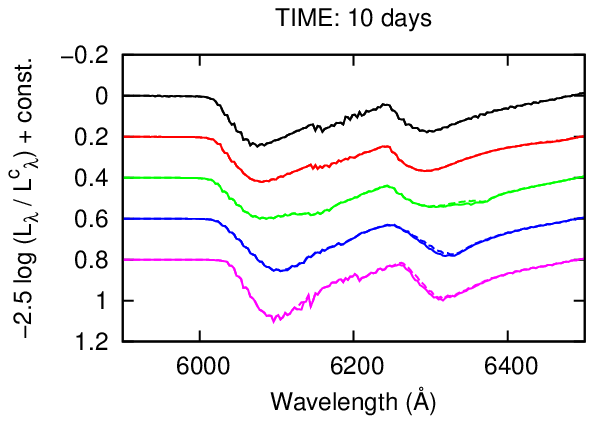}
  \end{minipage}
  \begin{minipage}{0.5\linewidth}
   \includegraphics[width=0.9\textwidth]{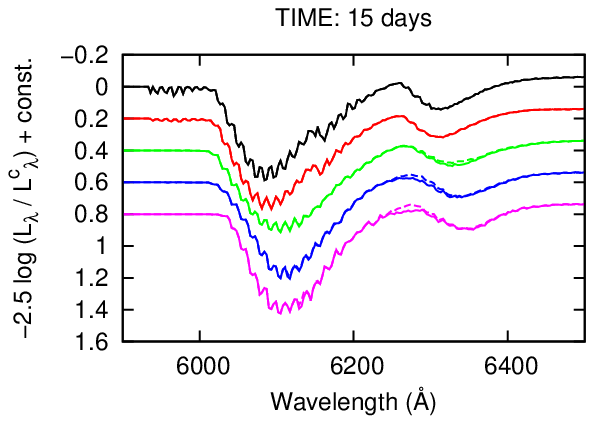}
  \end{minipage}
  \begin{minipage}{0.5\linewidth}
   \includegraphics[width=0.9\textwidth]{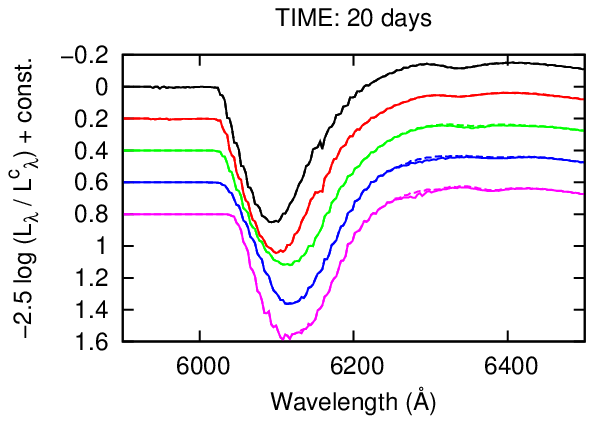}
  \end{minipage}
  \begin{minipage}{0.5\linewidth}
   \includegraphics[width=0.9\textwidth]{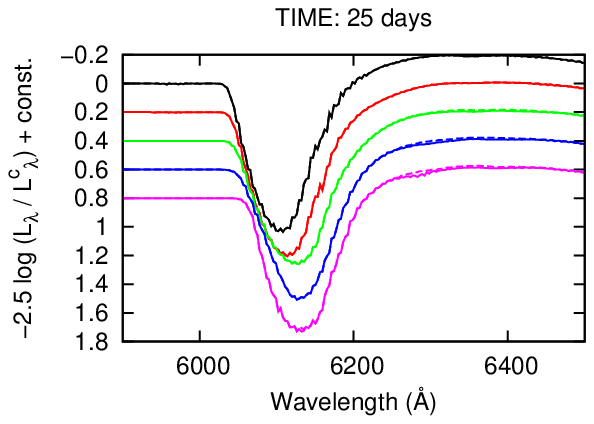}
  \end{minipage}
  \begin{minipage}{0.5\linewidth}
   \includegraphics[width=0.9\textwidth]{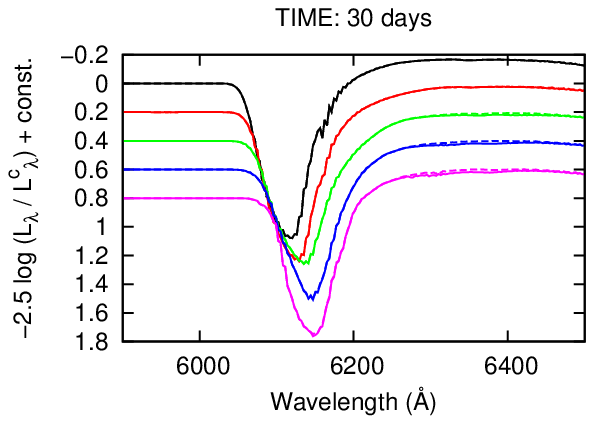}
  \end{minipage}
  \begin{minipage}{0.5\linewidth}
   \includegraphics[width=0.9\textwidth]{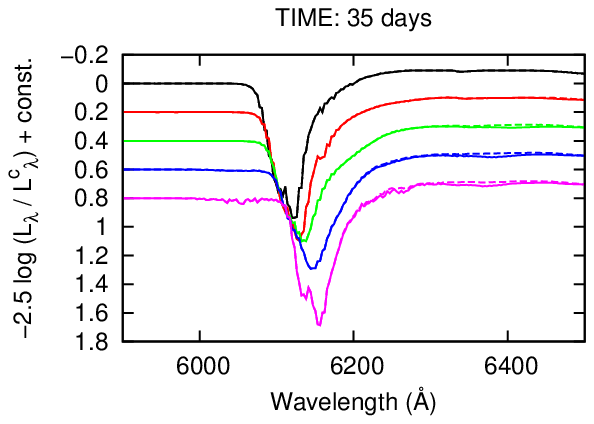}
  \end{minipage}
  \begin{minipage}{0.5\linewidth}
   \includegraphics[width=0.9\textwidth]{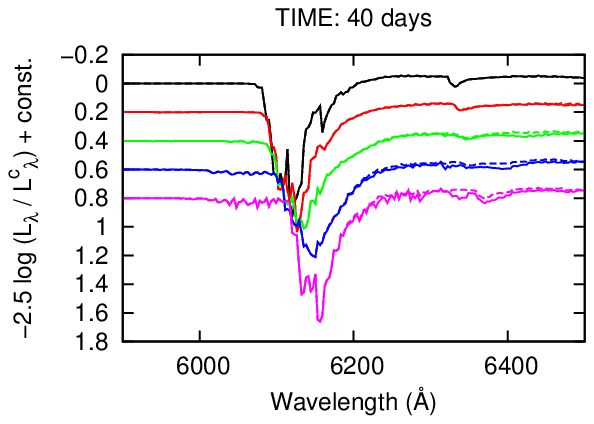}
  \end{minipage}
  \caption{Spectra in model RGb
  at day 5, 10, 15, 20, 25, 30, 35 and 40
  after the explosion.
  ---
  The black, red, green, blue, and magenta lines show light curves when we observe the SN with the viewing angle $\theta = 0$, $2\pi/3$, $5\pi/6$, $11\pi/12$, and $\pi$, respectively.
  The solid lines are calculated including all spectral lines.
  Besides, the dashed lines are calculated without H$\alpha$ lines.
  They are normalized by the continuum spectra.
  }
  \label{spectra-RGb-2.eps}
\end{figure*}

We calculate spectra with several viewing angles $\theta$ every 5 days up to 40 days since explosion.
Figures~\ref{spectra-MS-2.eps}-\ref{spectra-RGb-2.eps} show spectrum of each model.
Here the monochromatic luminosity $L_\lambda$ is normalized by the continuum luminosity $L_\lambda^c$.
In the early spectra,
strong absorption features come from C II line ($\sim 6300$ \AA) and Si II line ($\sim 6100$ \AA).
On the other hand, H$\alpha$ feature is quite weak or hidden by the C II absorption,
because W7 model retains a large amount of carbon in the outer ejecta.
In reality, C II absorption features are weak and rarely detected in observed SN Ia spectra \citep{2012ApJ...745...74F}.
We overestimate their absorption feature and therefore there is a possibility to detect  H$\alpha$ feature in some SNe Ia depending on the carbon content and the viewing angle.

\begin{figure*}
  \begin{center}
   \includegraphics[width=0.5\textwidth]{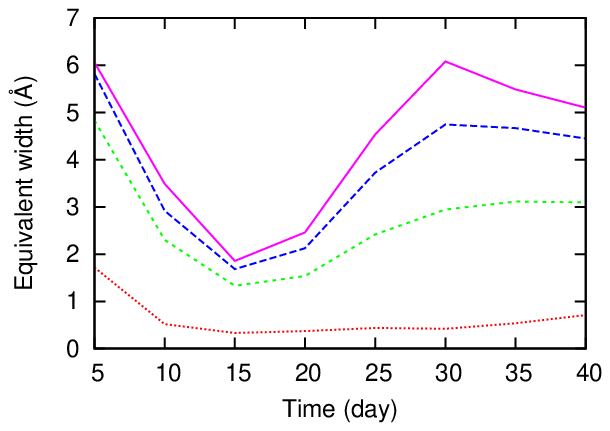}
   \includegraphics[width=0.5\textwidth]{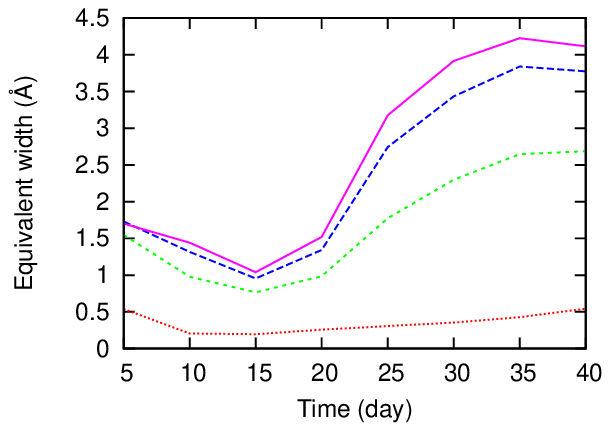}
   \includegraphics[width=0.5\textwidth]{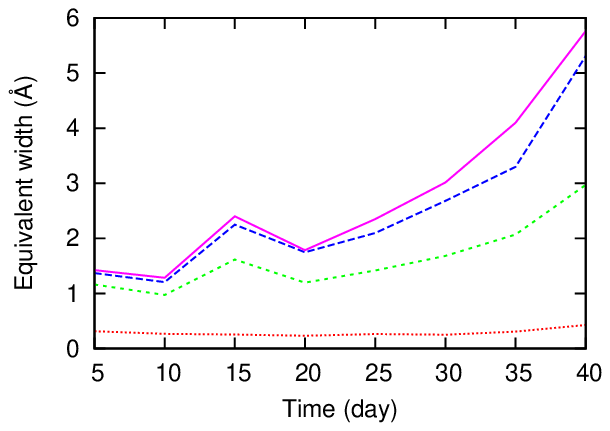}
  \end{center}
  \caption{Equivalent width of H$\alpha$ line in each model. ---
  The top panel shows model MS.
  The center panel shows model RGa.
  The bottom panel shows model RGb.
  The dotted, short-dashed, long-dashed, and solid lines show light curves when we observe the SN with the viewing angle $\theta = 2\pi/3$, $5\pi/6$, $11\pi/12$, and $\pi$, respectively.
  }
  \label{EW.eps}
\end{figure*}

Especially, model MS has a strong H$\alpha$ feature on its spectra before day 10 and after day 30.
As was mentioned before, the overestimate of the stripped mass in this model enhances the feature as compared with reality.
In models RGa and RGb, H$\alpha$ absorption becomes prominent only  after day 30. On the other hand, C II feature already becomes weak in this epoch
and does not disturb the H$\alpha$ feature.
The S/N ratio needed for the 3$\sigma$ detection of these features is written as
\begin{equation}
  S/N = 3 \times \frac{L_\lambda^c}{|L_\lambda - L_\lambda^c|}
\end{equation}
At day 35, the S/N ratio to detect the most strong feature in the spectrum is calculated as 59 for model RGa, and 125 for model RGb.
Figure~\ref{EW.eps} shows the time evolution of the equivalent width of H$\alpha$ line.
The peak value of the equivalent width with the viewing angle $\theta = \pi$ is
6\,\AA\ for model MS,
5\,\AA\ for model RGa,
and
4\,\AA\ for model RGb.
With $\theta = 5\pi/6$, the peak value decreases by a factor of about two compared with the above value in each model.
Thus, in models RGa and RGb, the H$\alpha$ feature can be detected when the viewing angle is in the range of $\theta > 11\pi/12$, or within the solid angle of $\Omega \sim 0.2$.
The event ratio in each model is estimated as $\Omega/4\pi \sim 0.02$.
However, this H$\alpha$ feature has never been observed in SN Ia spectra.
Thus, these systems are ruled out from the progenitor candidates.
Here, it should be noted that
these calculations assume that the ionization states are in thermal equilibrium and that we need to know the continuum spectra with sufficient accuracies in advance.
\citet{2014ApJ...794...37M} already discussed  hydrogen lines in this late phase in more detail including the Paschen lines.
%If we take into account the radiative excitation, the number of hydrogen atoms in excited states may increase.
%Then, H$\alpha$ absorption will become more easily detectable.

\begin{figure*}
  \begin{center}
   \includegraphics[width=0.5\textwidth]{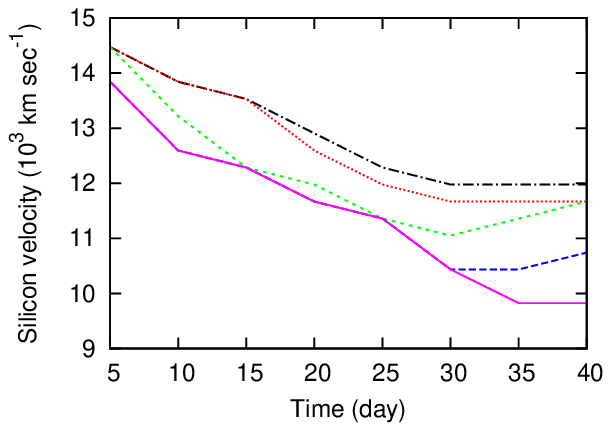}
   \includegraphics[width=0.5\textwidth]{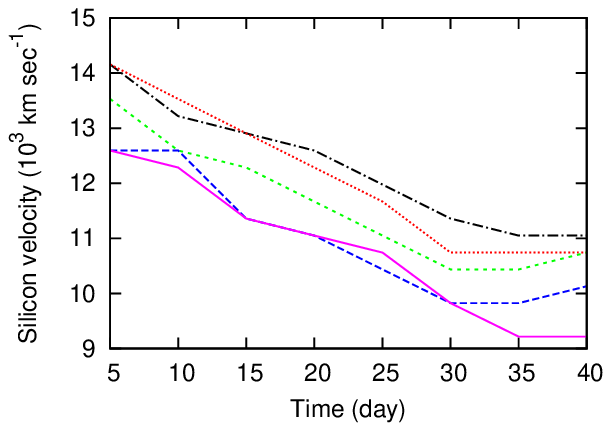}
   \includegraphics[width=0.5\textwidth]{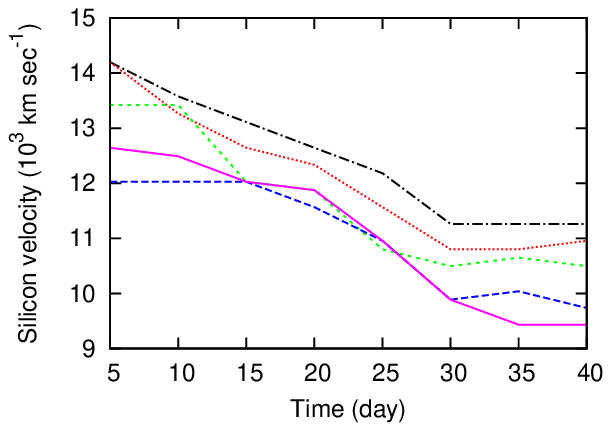}
  \end{center}
  \caption{Angular dependence on Si II line in each model. ---
  The top panel shows model MS.
  The center panel shows model RGa.
  The bottom panel shows model RGb.
  The dash-dotted, dotted, short-dashed, long-dashed, and solid lines show light curves when we observe the SN with the viewing angle $\theta = 0$, $\pi/2$, $2\pi/3$, $5\pi/6$, and $\pi$, respectively.
  }
  \label{vel-si.eps}
\end{figure*}

As another spectral feature to be investigated,
Figure~\ref{vel-si.eps} shows time evolution of silicon velocity with different viewing angles.
The differences of velocities between $\theta = 0$ and $\pi$ are about $1000\unit{km~s^{-1}}$ for all models. A spectral resolution of about 20\,\AA\ could distinguish the differences.
This variation occurs because the edge of the ejecta is decelerated and masked by the companion star.
We usually observe a SN from a single direction
and it is difficult to distinguish this effect from the variety of expansion speeds of individual SNe Ia.
One possibility to extract this effect is a SN Ia for which a several light echoes are discovered.
They will have different silicon velocities depending on the viewing angles, and can be compared with our results.
\citet{2008Natur.456..617K} obtained a light echo spectrum of SNR Tycho with the spectral resolution of 24\,\AA, observed by the FOCAS camera on the Subaru telescope.
The resolution we require is higher.
Furthermore, the light echo spectrum is constructed by a mixture of SN spectra for about 10 days.
Thus, it is not feasible to detect the variation at present.

These spectral features confirmed the results obtained by more sophisticated treatment of radiative transfer  in the same models  \citep{2014ApJ...794...37M} except that the temporal evolution of silicon line.  \citet{2014ApJ...794...37M}  obtained the silicon velocities independent of viewing angles at day $\sim10$ while we always see lower velocities when viewing from the companion side. This difference might be partly due to different treatments of the shock heating. \citet{2014ApJ...794...37M}  redetermined the temperature by imposing the radiative equilibrium condition ignoring the shock heating. Higher temperature due to the shock heating reduces the number of Si II ions and lowers the velocities at the absorption minima. In later phases, the adiabatic cooling diminishes  contributions of the shock heating and the results of the two different simulations converge.

\section{Summary and Conclusions}
In the SD scenario of SNe Ia, the collision between the ejecta and its companion occurs.
Due to this collision, their light curve and spectra are changed from the case of SN Ia without the companion.
These characteristics are expected to constrain progenitor systems.
In this study,
we perform radiation hydrodynamical simulations of this collision for three binary systems in the SD scenario.
Using the obtained snapshots, we solve the radiation transport equation,
and obtain their light curves and spectra with several viewing angles.
Then, we determine whether they can constrain the event rate of the SD progenitor.
The results are as follows.

%\begin{description}
%\item{Light curves:}
In comparison to \citet{2010ApJ...708.1025K},
the prompt emission due to the collision becomes weaker,
since the photospheric temperature becomes lower by the weaker coupling of gas and radiation.
In models MS and RGa, the luminosity does not exceed that of the $6M_\odot$ main sequence model in \citet{2010ApJ...708.1025K}, above which the emission can be detected in observed SNe Ia.
On the contrary, in model RGb, the luminosity still exceeds the threshold.
Thus, 
we conclude that
binary systems with the separation $A < 2\times 10^{13}\unit{cm}$
is allowed as a progenitor of SNe Ia.
As exemplified by SN 2014J, photometric observations during the first few days for  a large sample of SNe are crucial for distinguishing the two scenarios.

%\item{Spectra:}
In all models,
the equivalent widths of H$\alpha$ absorption become a few \AA\ 30 days after the explosion.
However, in model MS, the stripped mass is overestimated and the absorption will become weaker in reality.
On the other hand, in models RGa and RGb, the H$\alpha$ feature can be detected if the viewing angle $\theta > 11\pi/12$.
Then, 2\% of SNe Ia in these models should have the feature on their spectra,
while there is no sign of H$\alpha$ absorption in the observed spectra.
If a sufficient number of spectra were obtained with good S/N ratios
and were not showed this feature,
a model with a red giant companion filling the Roche lobe could be ruled out from SNe Ia progenitors.

The central wavelength of Si II absorption varies depending on viewing angles.
If more than one light echoes are discovered in a single SN, we obtain its spectra with different viewing angles.
Since the difference of silicon velocities depending on the viewing angles is about $1000\unit{km~s^{-1}}$,
the required resolution is about 20\,\AA.
The current observation has not reached this resolution for light echo spectra.
We need higher resolution which will be reached by future observations.

%\end{description}

As shown above, our models indicate that the prompt emission can not constrain binary systems with a short separation. One should note here that our models may significantly underestimate the luminosity due to the shock heating because we do not take into account the coupling of radiation and gas through Compton scattering. Thus we need to address the effects of Compton scattering before reaching a conclusion on this issue.
At the same time, we also found that the spectral feature might constrain the occurrence of thermalization  through Compton scattering.
Furthermore, H$\alpha$ feature brings us additional information of the progenitor of an individual SN Ia.
Because we can not predict where and when a SN will occur, the prompt emission is difficult to be found.
It will be detected only when a monitoring telescope is occasionally viewing around its site.
In contrast, some hydrogen features become prominent after the maximum light, which relaxes the required observing cadence.
If we continuously observe the SN and take spectra with a right viewing angle, we will never miss it.
This is a great advantage.

\subsection*{Fund}
This work was supported by JSPS KAKENHI Grant Number 23224004.

\newpage

% Figure 1

% Figure 2

% Figure 3

% Figure 4

% Figure 5

% Figure 6

% Figure 7

% Figure 8

% Figure 9

% Figure 10

% Figure 11

% Figure 12

% Figure 13

\end{document}